\documentclass[twocolumn,final,twoside,journal,twoside]{IEEEtran}\usepackage{lipsum}\usepackage{balance}

\usepackage{flushend}
\usepackage{soul}
\usepackage{graphicx,cite,amssymb,amsmath,color}
\usepackage[english]{babel}
\usepackage{calligra}
\usepackage[printonlyused]{acronym}
\usepackage[Glenn]{fncychap}
\usepackage{lettrine}
\usepackage{algorithm, algpseudocode}
\usepackage{cite}
\usepackage{filecontents}
\usepackage{amsmath}
\usepackage{bbm}
\usepackage{amsfonts}
\usepackage{fancyhdr}
\usepackage[font=scriptsize,labelfont=bf]{caption}
\usepackage{enumerate}
\usepackage{float}
\usepackage{mathrsfs}
\usepackage{caption}\usepackage{subcaption}
\usepackage{array}
\usepackage{placeins}
\usepackage{lineno}
\usepackage{soul}
\usepackage{mathtools}
\usepackage{nopageno}
\usepackage{lipsum}
\usepackage{mathtools}
\usepackage{cuted}
\usepackage{xcolor}
\usepackage{epstopdf}
\usepackage{flushend}
\IEEEoverridecommandlockouts

\begin{document}
\title{ Caching  in Heterogeneous Networks with Per-File Rate Constraints}
\author{Estefan\'ia Recayte, {\em Student Member, IEEE} and  Giuseppe Cocco,
{\em Member, IEEE}  
\thanks{
Estefan\'ia Recayte is    with the Institute of Communication and Navigation of the Deutsches Zentrum
f\"ur Luft- und Raumfahrt (DLR), 82234 Weßling, Germany. Email: estefania.recayte@dlr.de.   and with  DEIS, University of
Bologna, 40136 ${\text{Bologna, Italy.}}$ 

Giuseppe Cocco  is with the LTS4 Signal Processing Laboratory   and the Laboratory of Intelligent Systems (LIS) of the \'Ecole Polytechnique F\'ed\'erale de Lausanne (EPFL). Email: giuseppe.cocco@epfl.ch. 

Giuseppe Cocco is partly founded by the European Union's Horizon 2020 research and innovation programme under the Marie Sklodowska-Curie Individual Fellowship grant agreement No. 751062.
}
\thanks{This work has been accepted for publication at IEEE Transactions on Communications 2018.} \thanks{ \copyright 2018 IEEE. Personal use of this material is permitted. Permission
from IEEE must be obtained for all other uses, in any current or future media, including
reprinting /republishing this material for advertising or promotional purposes, creating new
collective works, for resale or redistribution to servers or lists, or reuse of any copyrighted
component of this work in other works.}
 }

\maketitle
\thispagestyle{plain}
\pagestyle{plain}
\vspace{-0.7in}
\begin{abstract}
 We study the problem of caching optimization in heterogeneous networks with mutual interference and per-file rate constraints from an energy efficiency perspective. A setup is considered in which two cache-enabled transmitter nodes and a coordinator node serve two users. 
 We analyse  and compare two approaches: (i) a cooperative approach    where each of the transmitters might serve either of the users and (ii) a non-cooperative approach in which each transmitter serves only the respective user. We formulate the  cache allocation optimization problem  so that the overall system power consumption is minimized while the use of the link from the master node to the end users is spared whenever possible.  We also propose a low-complexity optimization algorithm and show that it outperforms the considered benchmark strategies. 
 
  Our results indicate that significant gains both in terms of power saving  and sparing of master node's resources can be obtained when full cooperation between the transmitters is in place. Interestingly, we show that in some cases storing the most popular files is not the best solution from a power efficiency perspective. \end{abstract}
 %\vspace{-0.2in}
\vspace{-0.2in}
\section{Introduction}
Wireless data traffic has grown dramatically in recent years. It is foreseen that traffic demand in the fifth generation (5G) mobile communication networks will increase by 1000-fold by the year 2020 \cite{book5G}.  This huge demand will be mainly characterized by high-definition video  contents and streaming applications which not only require  a significant ammount of   bandwidth but also have stringent  quality of service requirements (QoS) \cite{book5G}.  In order to  offload the network and to reduce the system level energy consumption, \emph{cache-aided networks} have been  studied extensively in the recent years \cite{PhyCaching, DoFcaching2016, latency, 2layerCachin, cachesky, edgeEnergy, proactive, joint, proactive2, multicast, cached_inter_1, cached_inter_2, cached_inter_3, cached_inter_4, cached_inter_5, nsCaching, samesize, powerTransmission, UAV_zaho}. One of the main advantages
of caching is the possibility to spare backhaul resources by storing files during
periods in which the network traffic is low and directly serve
the users during high-load periods. 
 Due to the limited storage capacity of the transmitters, the   design of effective  caching policies is required. 
 %%%%%%%%%%%%%%%%%%%%%%%%%%%%%%%%%%%%%%%%%%%%%%%%

{ Cache-aided networks have been studied in literature with the aim of minimizing   backhaul congestion  } \cite{edgeEnergy} { and energy consumption} \cite{joint, proactive2, multicast} { or  maximazing the throughput} \cite{cached_inter_1}  \cite{cached_inter_2}. {The issue of interference in cache-aided networks has also been studied in many recent works } \cite{cached_inter_1,cached_inter_2,cached_inter_3,cached_inter_4,cached_inter_5}. {However, most works consider an approach in which files are partitioned into sub-files and the same transmission rate over the channel is used for each of them. }
  
   In this paper, we focus on  a full cooperative caching interference network. By jointly considering per file rate requirements and popularity ranking,  we   determine the   content placement and the delivery strategies in order to optimize the energy efficiency of the system.  
 \subsection{Related works} 
Caching optimization has been investigated in recent works for different kinds of heterogeneous networks. In \cite{2layerCachin} a two layer caching model that aims at minimizing the satellite bandwidth consumption by storing information both at ground stations and at the satellite is proposed.  In \cite{cachesky} the authors consider cache-enable unmanned aerial vehicles (UAV) and study the optimal location, cache content and user-UAV association  with the aim to maximize users' quality of experience while minimizing the transmit power. 
 Most of the proposed solutions exploit the statistics of
past users' file requests (often referred to as \emph{popularity ranking}) for filling up local caches with the most popular contents \cite{edgeEnergy}\cite{proactive}. 

Caching from an energy efficiency perspective has also been extensively studied. In \cite{multicast} the authors aim at reducing the total energy cost in a heterogeneous cellular network by applying a caching optimization algorithm for  multicast transmission.
  Energy minimization in the backhaul link is studied in \cite{joint}, where the authors propose a strategy for jointly optimizing  content placement and delivery, which leads to minimize  the energy consumption.  Energy efficiency is also studied in \cite{proactive2}, where  a proactive downloading strategy to exploit the good channel conditions and  minimize the total energy expenditure is proposed. 
 
Due to the significant network densification expected in 5G and beyond,  interference management represents a key challenge.  
 In order to counteract interference,  a combination of caching with  interference alignment (IA) and zero-forcing (ZF) techniques has been recently proposed. In their seminal work \cite{cached_inter_1}, Maddah-Ali and Niesen leverage on IA and ZF to show that significant gains in terms of load balancing and degrees of freedom can be obtained when caches at both the transmitter and the receiver side are used. 
  In  \cite{cached_inter_4} and \cite{cached_inter_5} the  authors extend the work in \cite{cached_inter_1} considering a delivery phase in which different sub-files are transmitted over different channel blocks. 
In \cite{cached_inter_3} authors propose a file placement strategy that allows to create cooperation opportunities between transmitters through interference cancellation and IA.

To the best of our knowledge,  energy optimization in cooperative interference networks with caching capabilities only at the transmitter assuming a one-shot delivery phase and   considering different per-file rate requirements and  different popularity rankings   has not yet been analysed in literature.
  Due to the one-shot delivery phase and to the different per-file rate requirements, 
the minimum sum-power needed to satisfy users requests depends on both the files' rates and the state of the links involved in the transmission. Since the requests are not known during the placement phase, a statistical optimization algorithm leveraging on the file popularity ranking has to be used to choose the cache allocation achieving the minimum power. Due to the discrete nature of the optimization along the cache allocation dimension and  the non convexity of the objective function   in the sum-power dimension, it is  challenging  to find a closed form solution or an algorithm with a complexity that scales well with the number of files.
%\vspace{-0.2in}

\subsection{Contributions}
We optimize the placement and delivery phase for minimizing the power consumption of a cache-enabled network under QoS requirements and in the presence of interference.
Our main contributions are the following:
\begin{itemize}
\item Unlike previous works, files are delivered over an interference channel and different PHY rate requirements per file have to be respected.  While caching over interference channels has  been formerly studied \cite{powerTransmission}, in previous works the rate constraints are relative to the user rather than to the file, i.e.,  each file has the same rate requirement while the per-user rate over the channel can differ, in that it accounts for information relative to possibly more than one file. Moreover, in \cite{powerTransmission} only the broadcast (BC) strategy is considered and, unlike   the present work, receivers are equipped with a cache, which allows them to efficiently perform interference cancellation. 
\item Unlike \cite{multicast, cached_inter_2, cached_inter_3}, we assume a one-shot delivery phase, in which transmission takes place in a single channel block and all selected files are transmitted at the same time. 
\item Depending on cache content and requests, transmitters can apply one of the following strategies: multiple input multiple output (MIMO) broadcast transmission, the Han-Kobayashi rate splitting approach \cite{kobayashi}, multiple input single output (MISO) transmission, broadcast  and multicast (MC) transmission. Up to our knowledge no other paper so far jointly considered this set of techniques in the context of caching. 
\item Extending our preliminary work \cite{nsCaching}, we establish a correspondence between the cache allocation/file request and the  strategy that is implemented using the set of cooperative techniques listed in the previous point and we  derive the minimum power cost needed for satisfying the rate requirements.
\item We propose an iterative algorithm for the Han-Kobayashi strategy that asymptotically converges to the minimum required sum-power and is significantly more efficient than an exhaustive search algorithm using the same granularity. 

\item We propose a low-complexity suboptimal algorithm  to solve the cache allocation problem. The algorithm outperforms benchmark policies based on highest popularity and highest rate and has a complexity which grows linearly with the number of files.
\item Our numerical results show that memorizing the most popular files is not always the most convenient strategy when transmitters have to respect per-file rate requirements. Furthermore, thanks to the considered cooperation strategies a significant saving of the coordinator node's resources can be achieved with respect to a non-cooperative system. 
%Interestingly, we also observe that the optimal cache allocation is robust to imperfect knowledge of channel statistics.
 \end{itemize}
\vspace{-.01in}
The remainder of the paper is organized as follows. In Section \ref{sec:sysmod} we introduce the system model. In Section \ref{sec:Transmissions} the  different transmission strategies are presented  and the relative  power optimization subproblems are addressed. The cache optimization  is tackled in Section \ref{sec:optprob}. \ref{sec:num_res} contains the numerical results while the conclusions are presented in Section \ref{sec:conclusion}.
 
\section{System Model}\label{sec:sysmod}
 
 \begin{figure}[t]
\centering
\includegraphics[width=0.4\textwidth]{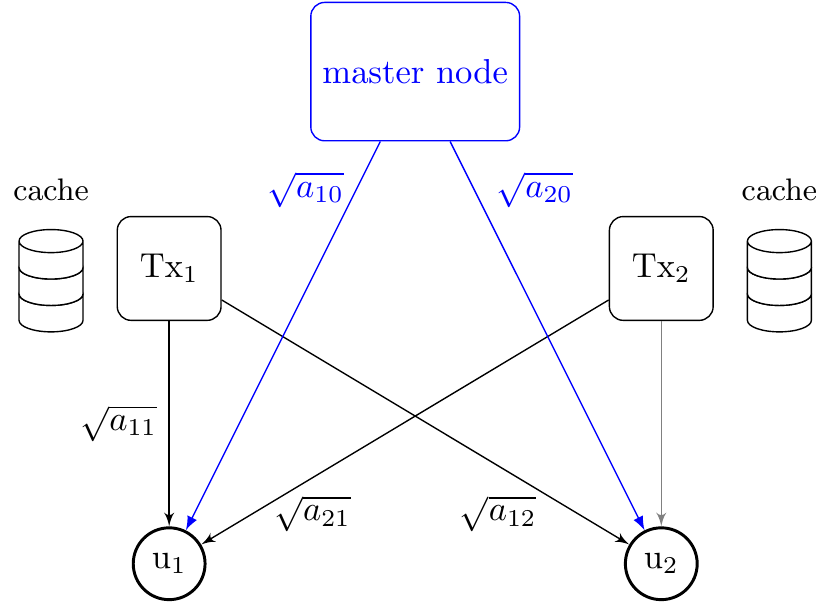}
\caption{System model of the considered heterogeneous network with caches at the transmitters.}
\label{fig:model}
\end{figure}
\subsection{Preliminaries}
We consider a heterogeneous network composed by a transmitter master node which has access to a collection of files (eg. video files), two transmitter nodes with limited caching capabilities and two users $u_n$, $n=\{1,2\}$. 
 All nodes in the network are equipped with a single antenna. The analysis  presented can be extended to a multi-antenna setup, in which case a lower power consumption can be achieved.

 The study carried out in this paper can be extended to the case of a generic number of SBSs and users. In doing so, the Han-Kobayashi scheme can be replaced with a transmission approach similar to that considered in \cite{multicap}.

 For ease of  exposition in the following  we assume a 5G network setup, with a macro base station (MBS) as master node  and  two  small base stations (SBS)  as transmitter nodes. However, the optimization problem and the transmission techniques presented in this work are not limited to a terrestrial architecture. In fact, in a heterogeneous satellite network a geostationary (GEO) satellite  connected to a ground station could act as the master node with access to all the files while low Earth orbit (LEO) satellites connected to the respective cache-enabled ground stations act as  transmitter nodes. Another possibility that is increasingly receiving  attention both from the research community and the industry is to employ unmanned aerial vehicles (UAV) as mobile base stations. In such context, a high altitude UAV, a satellite or an MBS could act as master node  while drones flying at lower altitude take the role of transmitters. 
  
  We assume that SBSs and MBS work in two separate frequency bands. The reason for choosing such setup is that it is likely to be implemented in the next generation of mobile networks (5G)  \cite{5gsbs}. In 5G, the SBSs will be assigned frequencies in the mmWave  that, thanks to the large bandwidth available, allow to reach high data rates, while the MBS operate at lower frequency, providing continuous coverage to mobile users with lower data rates. An orthogonal bandwidth setup is also foreseen in next generation heterogeneous satellite networks \cite{EDRS} and could be as well employed in UAV-assisted networks.

  We assume that user $u_n$  is associated to SBS$_n$.   
   Let us denote with $\sqrt{a_{n0}}$ the  channel coefficient between $u_n$ and the MBS while $\sqrt{a_{nm}}$ denotes the channel coefficient between $u_n$ and SBS$_m$, $m=\{1,2\}$, as shown in Fig.~\ref{fig:model}. We also define $a_{+0} \buildrel\triangle\over =\max\{a_{10}, a_{20}\}$ and  ${a_{-0}\buildrel\triangle\over=\min\{a_{10}, a_{20}\}}$ while $R_+$ ($R_-$) is the required rate of the file requested by the user with MBS channel coefficient  $a_{+0}$ ($a_{-0}$). Similarly $P_+$ and $ P_-$ are defined as the power used for files with rates $R_+$ and $R_-$, respectively.
 For SBSs we define ${a_{+1}\buildrel\triangle\over=\max\{a_{11}, a_{21}\}}$, ${a_{-1}\buildrel\triangle\over =\min\{a_{11}, a_{21}\}}$,  ${a_{+2}\buildrel\triangle\over=\max\{a_{22}, a_{12}\}}$, ${a_{-2}\buildrel\triangle\over=\min\{a_{22}, a_{12}\}}$ while $R_+$, $R_-$, $P_+$ and $P_-$ are defined for SBS$_1$ and SBS$_2$ similarly as for the MBS.
\begin{figure}[t]
    \centering
    \begin{subfigure}[b]{0.16\textwidth}
        \includegraphics[width=\textwidth]{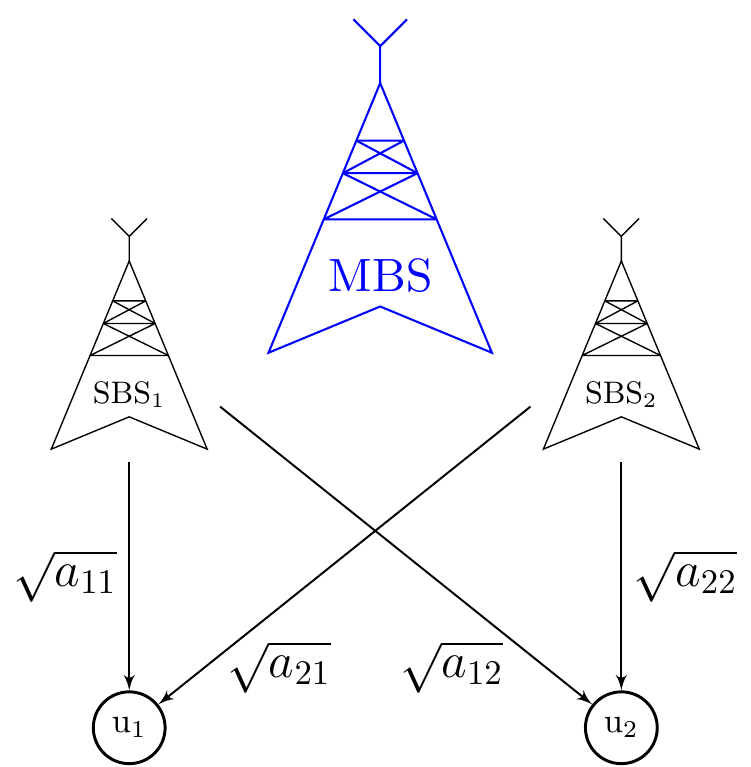}  \caption{}
    \end{subfigure} 
    \begin{subfigure}[b]{0.15\textwidth}
        \includegraphics[width=\textwidth]{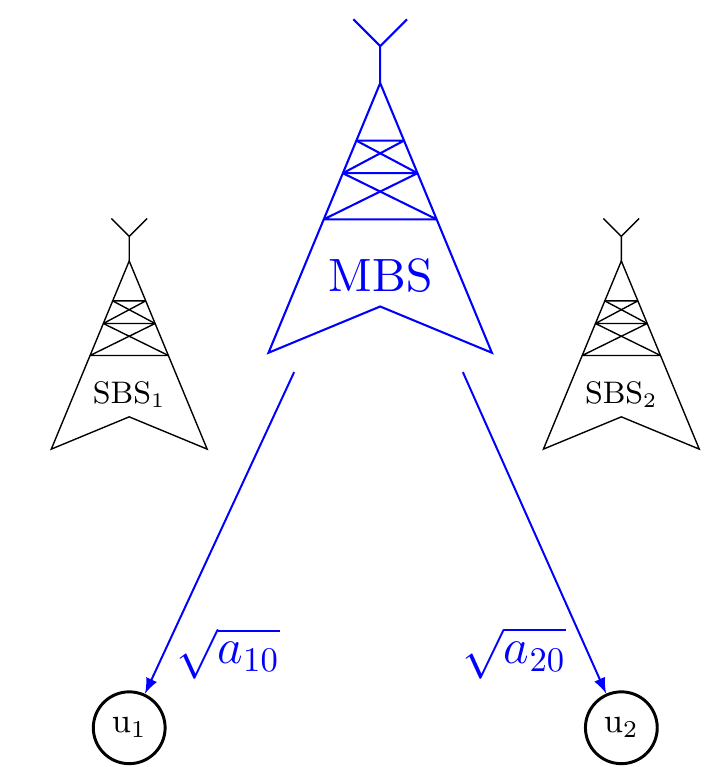}   \caption{}    
    \end{subfigure}
    \begin{subfigure}[b]{0.15 \textwidth}
        \includegraphics[width=\textwidth]{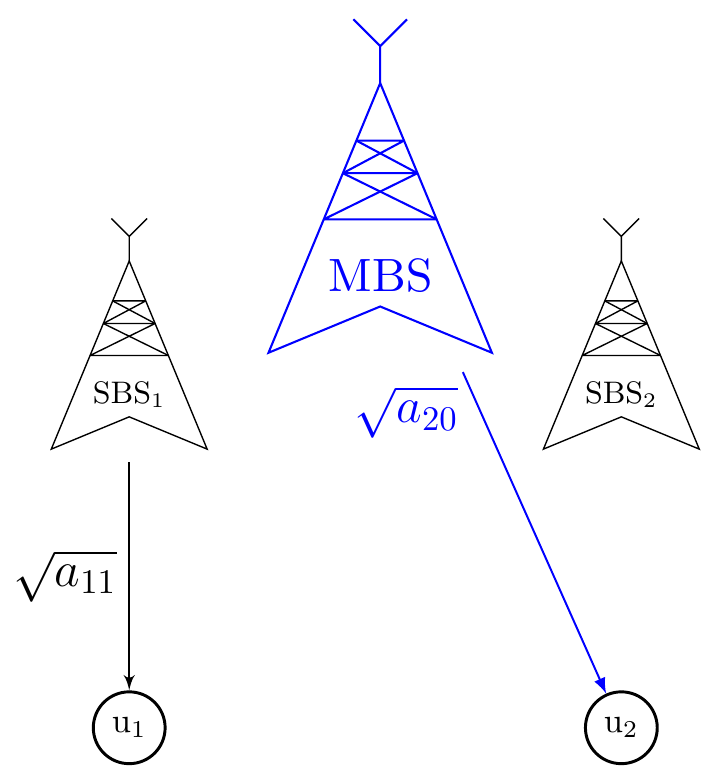}             \caption{}
  
     \end{subfigure}
    \caption{System model for the non cooperative approach: (a) interference as noise, (b) broadcast or multicast channel and (c) orthogonal channel}\label{fig:mod_nc}
\end{figure}

 The MBS has access to a library of $N$ files $\mathcal{F}$=\{$f_1$, \ldots, $f_N$\}. We assume that all files have the same size. This assumption is justifiable in practice since files can be divided into blocks of the same size \cite{samesize}. Each file $f_i$ has a minimum required transmission rate $R_i$, i.e., the minimum rate measured in bits/s/Hz at which the file has to be transmitted to the user in order to satisfy the QoS constraints\footnote{Although  we consider a constraint in terms of transmission rate over the channel, this can be easily translated into a constraint in terms of delay and quality, which is particularly suited to model video and image transmission.}.
The probability that $f_i$ is requested by a user is called \emph{popularity ranking}  and is indicated as $q_{i}$.
We assume that SBS$_n$ is equipped with a local cache of  size $M$ the content of which is denoted by $\mathcal{M}_n$.  

The binary variable $x_{in}$   indicates whether file $f_i$ is present in the memory of SBS$_n$ ($x_{in}=1$) or not ($x_{in}=0)$.  We further define  $\bar{x}_{in} \triangleq 1-x_{in}$.
 We denote with $\mathbf{x}_1$ and $\mathbf{x}_2$ the allocation vectors containing the values for $x_{i1}$ and $x_{i2}$, respectively. 
Depending on whether the files requested by the users are  present in the cache or not and on whether a cooperative scheme is considered or not, the system implements a different  transmission approach. The possible approaches  are explained in Section \ref{sec:Transmissions}. We indicate with $\mathbbm{1_{\alpha}}$ the indicator function which takes value $1$ if  ${\alpha}>0$ and $0$ otherwise. We further define  $\mathbbm{\bar{1}}_{\alpha}\triangleq 1- \mathbbm{1_{\alpha}}$.
We denote with $P_{i}^{(n)}$ the power used by the transmitter for sending $f_i$ to user $u_n$ at rate $R_i$. All transmission rates in the following are intended per complex dimension.  We denote with $\textbf{d}=[f_i \quad f_j]$   the users' demand vector. Without loss of generality we assume that $f_i$ is requested by $u_1$ and $f_j$ by $u_2$. 

 The value $c_{T}(i,j)$ indicates the minimum power consumed for serving the users when the system implements the transmission approach $T$. Such value is the  sum of two transmission powers  $c_{T}(i,j)=P_{i}^{(1)} + P_j^{(2)}$ if two transmitters are active while in case one transmitter takes care of both requests, the power cost is denoted as $c_{T}(i,j)=P_{i,j}^{(1,2)}$.  
 Finally, we indicate with  $Q_c(\mathbf{x}_1, \mathbf{x}_2)$ and $Q_{nc}(\mathbf{x}_1, \mathbf{x}_2)$ the expected power cost for serving a request  in the cooperative and in the non-cooperative  approach, respectively.

% \vspace{-0.3in}
\subsection{System Architecture}
 We assume that the SBSs are connected to the MBS through a backhaul link  while users  have a direct radio frequency link to both the SBSs and the MBS.   %The caching procedure consists of two phases: the placement phase and the delivery phase. 
 
 In the placement phase the SBSs  fill their own caches without deterministic knowledge of the user demands. Our aim is to understand which is the best strategy to choose the files for each transmitter's cache so that  the average power consumption is minimized during the delivery phase.
 
  In the delivery phase each user requests a file. We assume that the SBSs and the MBS can listen to the requests of both users and the communication is error-free\footnote{In practice this can be achieved by noting that the size of the request message is small and assuming a low channel code rate.}.  We assume that SBSs and MBS estimate the channel during the request  and  the channel does not vary during the file transmission while changes between two consecutive transmissions. Channel state information at the transmitter (CSIT) is assumed.  The  master node has knowledge of all the channel coefficients, the cache content and the user requests. According to this knowledge the MBS decides how the users have to be served and, eventually, coordinates the transmitters.
The file request and the file transmission take place in consecutive time slots and each user has to wait until the end of the transmission slot before placing a new request\footnote{Since files have the same size but different rate constraints, the transmission of the file with  highest rate terminates before the other file's. This implies an interference-free transmission during the part of the slot for the latter. In this paper we look for an upper bound on the minimum required power and assume that the interference is constant through the transmission}.
  
  Two different approaches are considered: a non-cooperative approach (NCA)  and a cooperative   approach (CA). In the NCA user $u_n$ is served by SBS$_n$ if the requested file  is present in the cache of SBS$_n$ and channel conditions (interference plus noise and fading levels) allow reliable communication at the required rate. If this is not possible, the user is served directly by the MBS. In the CA user $u_n$ can be served  by either the  SBSs or by the MBS if none of the SBSs  has the requested file. 
 The  CA  has two goals. On the one hand it aims at reducing the overall system's energy consumption and on the other hand it tries to spare the MBS   resources as much as possible. As a matter of facts in future heterogeneous terrestrial networks a single MBS   will serve a large number of SBSs \cite{5gsbs}.  
%%%%%%%%%%%%%%%%%%%%%%%%%%%%%%%%%%%%%%%%%%%%%%%%%%%%%%%%%%%%%%%%%%%%%%%%%%%%%%%%%%%%%%%%%%%%%%%%%%%%%%%%%%%%%%%%%%%%%%%%%%%%%%%%%%% 
 Thus, in order to avoid the MBS being a bottleneck, priority is always given to SBS transmissions whenever possible.
 
%%%%%%%%%%%%%%%%%%%%%%%%%%%%%%%%%%%%%%%%%%%%%%%%%%%%%%%%%%%%%%%%%%%%%%%%%%%%%%%%%
  \begin{figure}[t]
    \centering
    \begin{subfigure}[b]{0.16\textwidth}
        \includegraphics[width=\textwidth]{GIC}            \caption{}
    \end{subfigure}     
    \begin{subfigure}[b]{0.15 \textwidth}
        \includegraphics[width=\textwidth]{BC_m}              \caption{}
        \end{subfigure}    
     \begin{subfigure}[b]{0.15 \textwidth}
        \includegraphics[width=\textwidth]{ORT}            \caption{}
         \end{subfigure}      
      \begin{subfigure}[b]{0.15 \textwidth}
        \includegraphics[width=\textwidth]{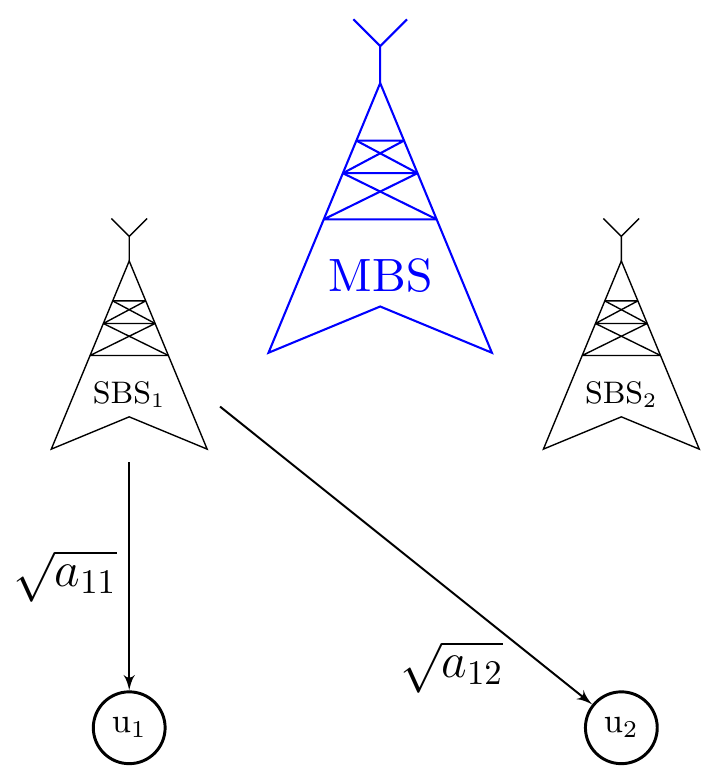}             \caption{}  
    \end{subfigure}    
     \begin{subfigure}[b]{0.15 \textwidth}
        \includegraphics[width=\textwidth]{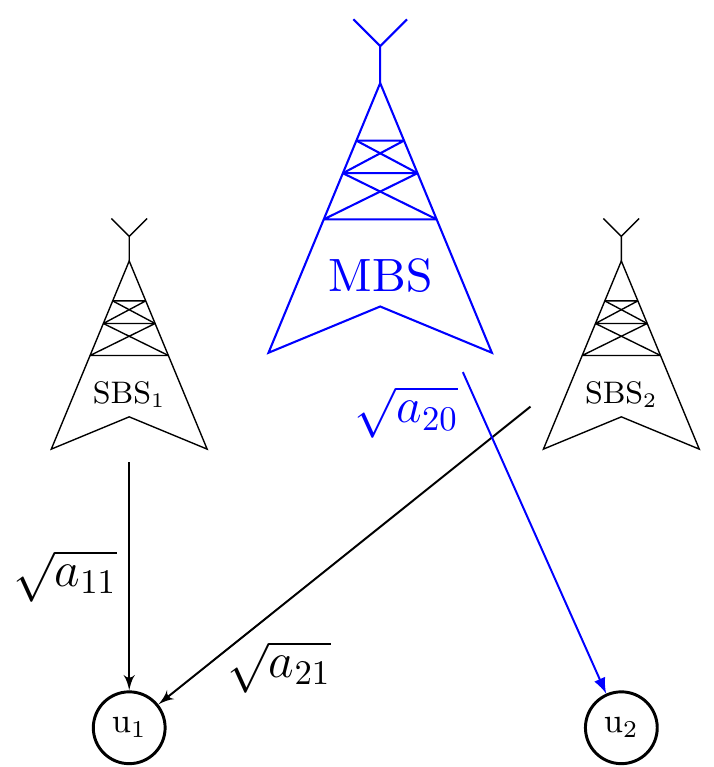}               \caption{}
      \end{subfigure}     
\caption{System model for  the cooperative approach: (a) interference channel without common information or  interference channel with common information or MIMO broadcast channel, (b) broadcast or multicast channel from MBS, (c) orthogonal channel, (d) multicast or broadcast channel from SBS and (e) MISO channel. }\label{fig:mod_coop}
\end{figure}
\vspace{-0.1in}
\section{Channel Models and Power Minimization Subproblems}\label{sec:Transmissions}
In this section we introduce the different transmission approaches with the relative channel models. For each of them we minimize the power  $c_{T}(i,j)$ required to transmit $f_i$ to $u_1$ and $f_j$ to $u_2$ at rate $R_i$ and $R_j$, respectively.  
 
In practical systems, a maximum per-node power constraint is present due to the physical limitations of the transmitters.  Unless otherwise stated, in the following we assume that such maximum power  is larger than that required for each considered setup. This allows us to get an insight on the power efficiency and MBS resource sparing, which is a step towards the full understanding of interference-limited caching systems. Note that using such assumption is not new in the literature related to caching. In \cite{powerTransmission}, for instance, bounds on the minimum average required power in a caching systems are studied with no explicit limits in terms of maximum per-node power.
  
The MBS decides which of the strategies described in the following has to be applied. The chosen strategy depends on which file has been requested and on the cache content of the SBSs. From now on, the subscripts in $P_i^{(1)}$ and $P_j^{(2)}$ are removed if there is no ambiguity.  
 
\subsection{ Gaussian Interference Channel with Common Information (GI$_c$)}
This channel is considered in the CA when the demand vector of the users is $\textbf{d}= [f_i \quad f_i]$  and $x_{i1} = 1$, ${x_{i2} = 1}$, (Fig. \ref{fig:mod_coop}-a).
We consider the interference channel model in standard form, which was introduced in \cite{carleial}. Starting from the physical model, the standard form is obtained as follows. Let us consider the physical channel model 
%\vspace{-5mm}
\begin{equation} \begin{aligned} 
\label{syst11}  y_1 &= \sqrt{a_{11}} \cdot x_1 + \sqrt{a_{12}}\cdot x_2 + z_1, \\
   y_2 &= \sqrt{a_{21}}\cdot x_1 + \sqrt{a_{22}} \cdot x_2 + z_2,   
\end{aligned}
\end{equation}
 where  $z_1 $ and $z_2$ are independent Gaussian variables with zero mean and unit variance. From the point of view of the achievable rates such model is equivalent to \cite{carleial} 
%\vspace{-5mm}
\begin{equation}
\begin{aligned}
    y_1 &=  x_1 + \sqrt{c_{12}}\cdot x_2 + z_1,\\
\label{syst22}   y_2 &= \sqrt{c_{21}}\cdot x_1 + x_2 +  z_2,
\end{aligned}
\end{equation}
where % $ a_{11}=1$, $a_{22}=1$,
  $c_{12} = a_{12}/a_{22}  $, $c_{21} = a_{21}/a_{11} $.
  Let us denote with $P^{(1)} $   ($P^{(2)}$) the physical power of $x_1$ ($x_2$). Given the power $\tilde{P}^{(n)}$ in the standard model, the physical power  is: $P^{(1)} = \tilde{P}^{(1)}/a_{11}$ and $P^{(2)} = \tilde{P}^{(2)}/a_{22}$.
In the GIc the same file $f_i$ has to be transmitted to both users at rate $R_i >0 $. Using the results in \cite{IC} and noting that in our case only a common message is to be transmitted, it can be easily shown that the power minimization problem can be written as 
\begin{equation*} 
\begin{aligned}
& \displaystyle \underset{P_i, P_j}   {\text{minimize }}  \displaystyle  & & \frac{P_i^{(1)}}{a_{11} } + \frac{P_j^{(2)}}{a_{22} }, \\
& \displaystyle \text{subject to } & & \displaystyle    \frac{1}{2}\log_2\left(1 + \left(\sqrt{P_i^{(1)}}+ \sqrt{c_{12} \cdot P_j^{(2)}}\right)^2\right)\geq   R_i, \\    
& \displaystyle   & & \frac{1}{2}\log_2\left(1 + \left(\sqrt{P_j^{(2)}}+ \sqrt{c_{21} \cdot P_i^{(1)}}\right)^2\right)\geq   R_i, \\
& \displaystyle  & & P_i^{(1)}, P_j^{(2)} \geq 0. 
\end{aligned}
\end{equation*}
The minimum power cost  $c_{GIc}(i,j)= P_i^{(1)} + P_j^{(2)}$ is derived in  Appendix I.
\vspace{-0.14in}
\subsection{Gaussian Interference as Noise (GIN)}
 
This channel  is considered in the NC approach when $x_{i1} = 1$, $x_{j2} = 1$ and $\mathbbm{1_{\alpha}}=1$ (Fig. \ref{fig:mod_nc}-a). 
where $\mathbbm{1_{\alpha}}$ is the indicator function 
%taking values $1$ if $\alpha >0$ and $0$ otherwise
and  $\alpha =  a_{22}  + \frac{a_{21} \cdot a_{12}}{a_{11} } \cdot (2^{2Rj}-1) \cdot (1-2^{2R_i})$. 

The channel model is given by   (\ref{syst11}).  
In this transmission scheme,  each SBS considers the interference generated by the other SBS as noise. The minimization problem is presented in \cite{nsCaching} and not reported here for a matter of space. 
 The minimum cost is achieved when \[P_i =   \frac{1}{a_{11} }\cdot  (2^{2R_i}-1) \cdot (a_{12} \cdot P_j +1)  \] and 
\begin{equation}
\begin{aligned}
\displaystyle P_j =& \frac{(2^{2R_j}-1) \cdot   \left[ \frac{a_{21}}{a_{11} } \cdot  (2^{2R_i}-1) +1 \right]}{a_{22}  + \frac{a_{21} \cdot a_{12}}{a_{11} } \cdot (2^{2R_j}-1) \cdot (1-2^{2R_i}) }. \label{PjIC}
\end{aligned}
\end{equation}
The value of $P_j$ in (\ref{PjIC}) is positive if and only if $\alpha>0$. In such case the  SBSs are able to successfully deliver their files to the respective users.  
On the contrary, if $\alpha \leq 0$ no   $P_i^{(1)}$ and  $P_j^{(2)}$ exist such that  the minimum required rate  can be achieved. In such case, the SBSs are not able to serve their users and the MBS will serve both users either applying a broadcast (if files are different)  or a multicast  transmission (if files are the same).
 The cost of the GIN channel is  
 \begin{equation}
 \begin{aligned}
 c_{GIN}(i,j)& =   \frac{(2^{2R_j}\tiny{-}1)     \left[ \frac{a_{21}}{a_{11} }    (2^{2R_i}\tiny{-}1) +1 \right]}{a_{22}  + \frac{a_{21}   a_{12}}{a_{11} }   (2^{2R_j}\tiny{-}1)   (1\tiny{-}2^{2R_i}) } + \\& \frac{(2^{2R_i}\tiny{-}1)  }{a_{11} }    \Bigg(   \frac{a_{12}(2^{2R_j}\tiny{-}1)     \left[ \frac{a_{21}}{a_{11} }    (2^{2R_i}\tiny{-}1) +1 \right]}{a_{22}  + \frac{a_{21}   a_{12}}{a_{11} }   (2^{2R_j}\tiny{-}1)   (1\tiny{-}2^{2R_i}) }  \tiny{+}1 \Bigg). \nonumber
 \end{aligned}
 \end{equation}
 \vspace{-.25in}
\subsection{Multicast Channel}
  
  This channel is considered when  $\textbf{d}= [f_i \quad f_i]$ in the following cases: (i) in both CA and NCA approaches when requests are satisfied by the MBS ($x_{i1} = 0$, $x_{i2}=0$), in which case, since both users require the same file, the MBS sends the file at the level of power needed to satisfy the $u_n$ experiencing the worst channel condition, (ii) in the CA when both requests are satisfied by one SBS     ($x_{i1} = 1$, $x_{i2} = 0$)  or  ($x_{i1} = 0$, $x_{i2} = 1$) and (iii) in  the NCA when  $\mathbbm{1_{\alpha}}=0$. The CA is represented in Fig.  \ref{fig:mod_coop}-b,c  while the NCA in Fig. \ref{fig:mod_nc}-a.
 
  For the case of multicast or broadcast transmission the channel model is 
\begin{equation}
 \begin{aligned}
   y_1 = \sqrt{a_{10}}\cdot x_1 + z_1, \\
   y_2 = \sqrt{a_{20}}\cdot x_2 + z_2.
\end{aligned}
\end{equation}
The minimum power required for $u_1$ is $P_i^{(1)}=  \frac{2^{2 R_i} -1}{a_{10}}$ while for $u_2$ is $P_i^{(2)} = \frac{2^{2 R_i} -1}{a_{20}}$. Thus, we have 
${P_i^{(1,2)} =  \max \left\lbrace \frac{2^{2 R_i} -1}{a_{10}},\frac{2^{2 R_i} -1}{a_{20}}\right\rbrace  =   \frac{2^{2 R_i} -1}{a_{-0}}}$.  
The cost of this channel is \[ c^{\text{\tiny MBS}}_{\text{\tiny MC}}(i,j) = \frac{2^{2 R_i} -1}{a_{-0}}. \] Similarly, for a multicast transmission from SBS$_n$ we have that $c^{\text{\tiny SBS}}_{\text{\tiny MC}}(i,j) = \frac{2^{2 R_i} -1}{a_{-n}}$.
\vspace{-.1in}
  \subsection{Broadcast Channel}
  \vspace{-0.05in}
In the broadcast channel, we have one transmitter and two receivers which request different files \cite{bookCoverThomas}.  This  type of channel is implemented by the MBS or one of the SBSs   when $\textbf{d}= [f_i \quad f_j],$  $i \neq j$, in the following cases: (i) in the CA when files are not present in SBSs caches  ($x_{i1} = 0$, $x_{i2} = 0$, $x_{j1} = 0$, $x_{j2} = 0$), in this case the MBS serves both users, (ii) in the CA  when one SBS   has both files while the other has none of them, that is when  ($x_{i1} = 1$, $x_{i2} = 1$, $x_{j1} = 0$, $x_{j2} = 0$) or ($x_{i1} = 0$, $x_{i2} = 0$, $x_{j1} = 1$, $x_{j2} = 1$), in this case a SBS serves both users and (iii)   in NCA when the file requested by $u_n$ is not present in $SBS_n$ for $n=1,2$ thus the MBS serves both users ($x_{i1} = 0$, $x_{j2} = 0$). Finally, (iv) the broadcast channel is implemented  in the NCA when $\mathbbm{1_{\alpha}}=0$.  The CA is represented in Fig.  \ref{fig:mod_coop}-b,c  while the NCA in Fig. \ref{fig:mod_nc}-a.

The power minimization problem to be solved, considering an MBS transmission, can be found in \cite{nsCaching} and thus is not reported here since it is formally the same as in case of SBS transmission.
Let $P_{-}, P_{+}$ be the powers related to the worst and best channel, respectively, as defined in Section II-$A$. The overall cost of this channel  is   $$c^{\text{\tiny MBS}}_{\text{\tiny BC}}(i,j) = \frac{2^{2R_+} - 1}{a_{+0}} + \frac{(2^{2R_-} -1)  (1+a_{-0} P_+)}{a_{-0}}.$$ Similarly, for the broadcast transmission from  SBS$_n$ we have: $$c^{\text{\tiny SBS}}_{\text{\tiny BC}}(i,j) = \frac{2^{2R_+} - 1}{a_{+n}} + \frac{(2^{2R_-} -1)  (1+a_{-n} P_+)}{a_{-n}}.$$
 
 \subsection{MIMO Broadcast}
  This channel is considered in the CA  when $\textbf{d}= [f_i \quad f_j]$, for $i \neq j$, and each SBS has both requested files  ($x_{i1} = 1, x_{i2} = 1, x_{j1} = 1, x_{j2} = 1$), as depicted in Fig.\ref{fig:mod_coop}-a.
 In this case, the MBS coordinates the SBSs transmission and provides the information needed (e.g. the  channel matrices) such that each SBS  calculates the portion of the MIMO codeword to be transmitted. 
 This channel is not considered in \cite{nsCaching} and is analyzed in the following. 
 
 We denote with $\mathbf{H}_1=[\sqrt{a_{11}}\ \sqrt{a_{21}}]^{T}$ and  $\mathbf{H}_2=[\sqrt{a_{12}}\ \sqrt{a_{22}}]^{T}$   the channel matrices between each SBSs   and the two users. 
 
In the MIMO BC channel the two single-antenna SBSs  collaborate for transmitting both files acting as a single multi-antenna terminal.
The dirty paper coding (DPC) approach is used. DPC consists in a layered precoding of the data so that the effect of the self-induced interference can be cancelled out \cite{DPC}. User messages are sequentially encoded. Let assume that the message for $u_1$ is encoded first. The decoding is such that a message experiences interference only from messages encoded after him. In this case, the message for $u_1$ is interfered by the message for $u_2$ %since the latter is encoded last, while 
and $u_2$ experiences no interference.
 
Let us define $\pi_m$ with $m=\{1,2\}$ as  a permutation describing the possible encoding orders such that  $\pi_1=\{2,1\}$ and $\pi_2=\{1,2\}$ and let $\pi_m(l)$ indicate  the $l$-th element of the encoding order $m$, e.g. $\pi_1(1)=2$ and $\pi_1(2)=1$.
The minimization problem to be solved is  
\begingroup
\allowdisplaybreaks
\begin{align} \label{BCopt}
%\begin{aligned}
&\underset{S_1, S_2, \pi_m} {\text{minimize}} & &\mathrm{tr} \left(\sum_{j=1}^2 \mathbf{S}_j\right) \\
&\text{subject to}&  &\frac{1}{2} \log_2 |\mathbf{I}  +  \widetilde{\mathbf{H}}_{\pi_m(2)}^T \mathbf{S}_{\pi_m(2)} \widetilde{\mathbf{H}}_{\pi_m(2)}| = R_{\pi_m(2)},  \nonumber \\ \nonumber
&&  &\frac{1}{2} \log_2 \frac{|\mathbf{I} + \sum_{j=1}^n\widetilde{\mathbf{H}}_{\pi_m(1)} \mathbf{S}_j \widetilde{H}_{\pi_m(1)} |}{|\mathbf{I}  +  \widetilde{\mathbf{H}}_{\pi_m(1)}^T \mathbf{S}_{\pi_m(2)} \widetilde{\mathbf{H}}_{\pi_m(1)}|} = R_{\pi_m(1)}, \\ \nonumber 
& & & m=1,2 
%\end{aligned}
\end{align}
\endgroup
where $\widetilde{\mathbf{H}}_n$ is a  square channel matrix obtained by opportunely modifying $\mathbf{H}_n$. Such modification is done as suggested in \cite{cioffi}, by introducing a virtual receive antenna at each user with virtual channel gain equal to $\epsilon$, with $\epsilon$ much smaller than any of the channel coefficients.  Thus,  $\widetilde{\mathbf{H}}_1$ = $\bigl(\begin{smallmatrix}
\sqrt{a_{11}} & \epsilon\\ \sqrt{a_{21}} & \epsilon
\end{smallmatrix} \bigr)$ and  ${\widetilde{\mathbf{H}}_2 =\bigl(\begin{smallmatrix}
\sqrt{a_{12}} & \epsilon\\ \sqrt{a_{22} } & \epsilon
\end{smallmatrix} \bigr)}$.  $\mathbf{S}_1$ and $\mathbf{S}_2$ are the  positive semi-definite input covariance matrices for $u_1$ and $u_2$ respectively while $\mathbf{I}$ is the noise covariance matrix. 
To solve (\ref{BCopt})  we transform the BC optimization problem into a  dual MAC optimization problem \cite{BCtoMAC}.
Then, the equivalent convex optimization problem can be written as  
%\vspace{-.2cm} \newpage
\begingroup
\allowdisplaybreaks 
\begin{align} \label{MACopt}
& \underset{\mathbf{S}_1^M, \mathbf{S}_2^M, \pi_m} {\text{minimize}} & &\mathrm{tr}\left(\sum_{j=1}^2 \mathbf{S}_j^M\right) \\ 
&\text{subject to}&  & \frac{1}{2} \log_2 | \mathbf{I} + \sum_{j=1}^2 \widetilde{\mathbf{H}}_j\mathbf{S}_j^M \widetilde{\mathbf{H}}_j^T|  = \bar{R}^M_{\pi_m(2)}, \nonumber \\ 
& &  & \frac{1}{2} \log_2 |\mathbf{ I} + \widetilde{\mathbf{H}}_{\pi_m(1)}\mathbf{S}_{\pi_m(1)}^M \widetilde{\mathbf{H}}_{\pi_m(1)}^T|  = \bar{R}^M_{\pi_m(1)},\nonumber \\ 
& && m = 1,2, \nonumber
\end{align} 
\endgroup
where $\mathbf{S}_j^M$ indicates the dual MAC covariance matrix. The following holds: $\text{tr}\left(\sum_{j=1}^2 \mathbf{S}_j\right)=\text{tr}\left(\sum_{j=1}^2 \mathbf{S}_j^M\right)$ and
${\bar{R}^M_i= \sum_{j=i}^2 R_i}$.
It is not straightforward to find the solution to problem (\ref{MACopt}). A suboptimal solution $\left(\mathbf{S}_1^M, \mathbf{S}_2^M\right)$ can be obtained using the iterative algorithm suggested in \cite{cioffi}. The minimum power is readily obtained as  ${P_i+P_j = \text{tr}\left(\mathbf{S}_1+\mathbf{S}_2\right)=\text{tr}\left(\mathbf{S}_1^M+\mathbf{S}_2^M\right)}$ and the overall cost of the channel is $c_{\text{\tiny MIMO}} (i,j)= P_i+P_j $.
 
\vspace{-0.15in}
\subsection{Orthogonal Channel}
\vspace{-0.05in}
This channel is considered  when $u_1$'s  ($u_2$'s) request is satisfied by the SBS and $u_2$'s ($u_1$'s) request is satisfied by the MBS using orthogonal channels. In the CA this occurs when $({x_{i1} = 1},$ ${x_{i2} = 0}, {x_{j1} = 0}, {x_{j2} = 0})$, ${(x_{i1} = 0}, {x_{i2} = 0}, {x_{j1} = 0}, {x_{j2} = 1})$ or $({x_{i1} = 0}, {x_{i2} = 1},$ ${x_{j1} = 0},$ ${x_{j2} = 0)}$, $(x_{i1} = 0, x_{i2} = 0, x_{j1} = 1, x_{j2} = 0)$. In the NCA such channel is considered when ${(x_{i1} = 1}, {x_{j2} = 0})$ or $({x_{i1} = 0},{ x_{j2} = 1})$ (Fig. \ref{fig:mod_nc}-c and Fig. \ref{fig:mod_coop}-c). Let us assume, without loss of generality, that $u_1$ is served by  SBS$_1$   and $u_2$ by the MBS. The channel model is the following 
\begin{equation*}
\begin{aligned}
   y_1 = \sqrt{a_{11}}\cdot x_1 +  z_1, \\
   y_2 = \sqrt{a_{20}} \cdot x_2 +  z_2.
\end{aligned}
\end{equation*}
%The power is minimized putting $P_i=\frac{2^{2R_i}-1}{a_{11} }$ and  $P_j=\frac{2^{2R_j}-1}{a_{20}}$. 
The cost of this channel is
\begin{equation*}
\begin{aligned}
 c_{\perp}(i,j) = \frac{2^{2R_i}-1}{a_{11} } +\frac{2^{2R_j}-1}{a_{20}}.
\end{aligned}
\end{equation*}
%\vspace{-0.1in}\vspace{-0.3in}
 \subsection{MISO Orthogonal}
 The multiple input single output orthogonal channel is considered in the CA when both SBSs have the file requested by one user but do not have the file requested by the other user $({x_{i1} = 1}, {x_{i2} = 1}, {x_{j1} = 0},{x_{j2} = 0}$ and ${x_{i1} = 0}, {x_{i2}=0}, {x_{j1} = 1}, {x_{j2} = 1})$ (Fig.\ref{fig:mod_coop}-e).
Without loss of generality, let us assume that $u_1$ is served by the two SBSs and $u_2$ is served by the MBS. We recall that all nodes have  a single antenna but, thanks to the coordination of the MBS, the two transmitters act as a multi-antenna terminal. The channel model is the following 
\begin{align*}
   y_1 &= \sqrt{a_{11}}\cdot x_1 + \sqrt{a_{12}} \cdot x_1 +  z_1,\\
   y_2 &=   \sqrt{a_{20}}\cdot x_2 + z_2 .
\end{align*}
The  problem to be solved is \cite{MISO} 
\begin{equation*}
\begin{aligned}
&\displaystyle  \underset{ P_i, P_j}   {\text{minimize}} & & \displaystyle   P_i + P_j,  \\
&\displaystyle \text{subject to } &  & \displaystyle    \frac{1}{2}\log_2\left( 1+ (a_{11} +a_{12}) \cdot P_i \right)\geq   R_i, \\
& \displaystyle & &   \frac{1}{2}\log_2\left(1+ a_{20} \cdot P_j\right)\geq   R_j, \\
& \displaystyle   & & P_i, P_j\geq 0,
\end{aligned}
\end{equation*}
where the solution is $P_i= \frac{2^{2R_i}-1}{a_{11}+a_{12} }$, $P_j= \frac{2^{2R_j}-1}{a_{20}}$ and the cost     is \[c_{\text{\tiny MISO}} (i,j)=\frac{2^{2R_i}-1}{a_{11} +a_{12}}+ \frac{2^{2R_j}-1}{a_{20}}.\]
 \subsection{Gaussian Interference Channel without Common Information (GI$_{wc}$)}
  In this channel the SBSs implement the Han-Kobayashi rate-splitting strategy\cite{kobayashi}. Such cooperative scheme consists in splitting each message rate into two parts. One part of the message is \emph{private} and is decoded only by the intended user while the other part of the message is \emph{common} and is decoded by both receivers. The sum of the rates of such parts is equal to the rate of the original message. Decoding part of the interfering signal allows to expand the capacity region with respect to the GIN case.
  This scheme is considered when  $\textbf{d}= [f_i \quad f_j]$, with $i\neq j$,   and requests are satisfied by the two SBSs ($x_{i1} = 1$, $x_{j2} = 1$, $x_{i2} = 0$, $x_{j1} = 0$) and  ($x_{j1} = 1$, $x_{i2} = 1$, $x_{i1} = 0$, $x_{j2} = 0$). In the latter case, each SBS  does not have the file requested from the corresponding user but has the file requested by the other user. Thus 
  SBS$_1$ serves $u_2$ and SBS$_2$ serves $u_1$.  (Fig. \ref{fig:mod_coop}-a). 
 
\begingroup
 \allowdisplaybreaks
\begin{algorithm}
 \caption{\small $c_{GIwc}=$ calculate\_Min\_Ptot\_HK($\Delta P^{\text{tot}}_{\text{min}}$,  $\Delta P^{\text{tot}}_{\text{max}}$, $\Delta P_{\text{min}}$, $\Delta P_{\text{max}}$, $\Delta \lambda_{\text{min}}$, $\Delta \lambda_{\text{max}}$, $ R_i, R_j$)}\label{algo1}
  \begin{algorithmic}[1]
 \small
 \State $c_{GIwc} = -1$; $P_{\text{tot}}=\Delta P^{\text{tot}}_{\text{max}}$; $\Delta P_{\text{tot}}=\Delta P^{\text{tot}}_{\text{max}}$;
  \While{ $\Delta P^{\text{tot}}>\Delta P^{\text{tot}}_{\text{min}}$}\Comment{Go on until the minimum granularity for $\Delta P^{\text{tot}}$ is reached}
  \State flag = 0  
  \State $\Delta P=\Delta P_{\text{max}}$; $\Delta \lambda=\Delta \lambda_{\text{max}}$ \Comment{Reset $\Delta P$ and $\Delta \lambda$}
  \While{flag == 0}
  \State create $G(\Delta P, \Delta \lambda)$ \Comment{Create search grid with current granularity excluding points previously checked}
  \For {$g \in G(\Delta P, \Delta \lambda)$ } \Comment{Each point $g$ is defined by coordinates ($P_i$, $P_j$, $\lambda_1$, $\lambda_2$)}
  \If {$g$ allows to achieve the pair ($R_i, R_j$)} 
  \State flag = 1
  \State $c_{GIwc} = P^{\text{tot}}$
  \State $\Delta P^\text{tot} = \Delta P^\text{tot}/2$ 
  \State $P^{\text{tot}} = P^{\text{tot}} - \Delta P^\text{tot}$ \Comment{Check if ($R_i, R_j$) are achievable with a smaller $P^{\text{tot}}$}
  \State \textbf{break for}
  \EndIf 
  \EndFor 
   \If {(flag == 0) }
    \If {($\Delta P> \Delta P_{\text{min}}$)}
    \State $\Delta P = \Delta P/2$ 
    \ElsIf {($\Delta \lambda> \Delta \lambda_{\text{min}}$)}
  \State $\Delta \lambda = \Delta \lambda/2$
  \ElsIf {$c_{GIwc} \neq  -1$} \Comment{Check if so far ($R_i$, $R_j$) has been achieved}
    \State $\Delta P^\text{tot} = \Delta P^\text{tot}/2$
    \State $P^{\text{tot}}=P^{\text{tot}}+\Delta P^{\text{tot}}$
  \Else
  \State $P^{\text{tot}}=P^{\text{tot}}+\Delta P^{\text{tot}}$\Comment{If ($R_i$, $R_j$) is not achievable with $P^{\text{tot}}$, the same holds for a smaller $P^{\text{tot}}$}
  \State \textbf{break inner while}
  \EndIf
  \EndIf
  \EndWhile
  \EndWhile
 \State \textbf{return} $c_{GIwc}$ 
  \end{algorithmic} 
 \end{algorithm}  \endgroup
  
Solving analytically the power minimization subproblem
for the case of GIwc is a challenging task due to the non convexity of the problem and  to
the intricacy of the conditions (see expressions (\ref{HK18})-(\ref{HK22})
and Appendix II). In order to calculate the cost $c_{GIwc} = P_i+P_j$, we developed the algorithm calculate\_Min\_Ptot\_HK, shown in Algorithm \ref{algo1} in pseudocode.

  Let us indicate with  $\Delta P^{\text{tot}}$   the granularity at which the sum power is evaluated, with  $\Delta P$  the granularity with which the sum power is divided between the SBSs $P_i$ and $P_j$ and with  $\Delta \lambda$  the granularity in which power is split between the private and the common message.\footnote{The three parameters are bounded as follows: $\Delta P^{\text{tot}}>0$, $0<\Delta P<1$ and $0<\Delta \lambda<1$. 
The algorithm takes as input the minimum and maximum granularities of $\Delta P^{\text{tot}}$, $\Delta P$ and $\Delta \lambda$ as well as the required file rates $R_i$ and $R_j$ and returns the minimum value of $c_{GIwc}$ that satisfies the rate constraints for the two cached files.} At initialization the algorithm performs the search over a grid of points\footnote{As an example, the first point of the grid for a given $P^{\text{tot}}$ is ($P_1$, $P_2$, $\lambda_1$, $\lambda_2$) = ($\Delta P      \cdot  P^{\text{tot}}$, $(1-\Delta P)P^{\text{tot}} $, $\Delta \lambda$, $\Delta \lambda$), the second point is ($P_1$, $P_2$, $\lambda_1$, $\lambda_2$) = ($2\Delta P  \cdot P^{\text{tot}}$, $ (1-2\Delta P)P^{\text{tot}}$, $\Delta \lambda$, $\Delta \lambda$) and so on.} spaced apart by the largest granularity  specified in the input in order to speed-up convergence. The granularity is progressively decreased  to the minimum value specified in the input to enhance  the accuracy of the solution.
 Note that the minimum power $c_{GIwc}$ is lower bounded  by $c_{\perp}$ and  upper bounded by $c_{GIN}(i,j)$.

The algorithm is much more efficient than exhaustive search over a grid with same minimum granularity. Specifically, if $\Delta P^{\text{tot}}_{\text{min}}\ll \Delta P^{\text{tot}}_{\text{max}}$, the gain in terms of number of operations is on the order of $\left(\Delta P^{\text{tot}}_{\text{max}}/\Delta P^{\text{tot}}_{\text{min}}-1\right)\times 1/\Delta \lambda_{\text{min}}\times 1/\Delta P_{\text{min}}$. The algorithm is based on the fact that, if a given sum power $P^{\text{tot}}$ does not allow to satisfy the rate constraints for any  rate- and power-split, then no $\tilde{P}^{\text{tot}}<P^{\text{tot}}$ can satisfy such constraints. This can be derived from the conditions in Appendix II. The algorithm converges to the absolute minimum asymptotically as $\Delta P^{\text{tot}}_{\text{min}}\rightarrow 0$, $\Delta \lambda_{\text{min}}\rightarrow 0$, $\Delta P_{\text{min}}\rightarrow 0$.
 Note that from   conditions  in Appendix II   follows that   a solution to the minimization problem always exist. Therefore, there is no need to check for the existence of a solution as done for the GIN channel in the NCA, where the indicator function $\mathbbm{1}_{\alpha} >0$  was required. 
 The proof is not reported here for a matter of space.
 
\begin{figure*} [h!]
 \begin{align}\label{Qxc}
   Q_\text{c}(\mathbf{x}_1, \mathbf{x}_2) & \tiny{=}  
  \sum_{f_i \in \mathcal{F}} \sum_{\substack{f_j \in \mathcal{F}  \\  \nonumber
   f_i\neq f_j}} q_{i}  q_{j}   \Big\{  c_{\text{GI}_{\text{c}}} (i,j)  \Big[  \sum_{t=i,j}\sum_{n=1,2} x_{tn} x_{\bar{t}n} x_{t\bar{n}} \bar{x}_{\bar{t}\bar{n}} \Big]      \tiny{+} c_{\text{  BC}}^{\text{\tiny SBS}}(i,j)   \Big[\sum_{n=1,2}  x_{in}   x_{jn}  \bar{x}_{i\bar{n}}   \bar{x}_{j\bar{n}} \Big]   \tiny{+} c_{\text{  BC}}^{\text{\tiny MBS}}(i,j)   \Big[  \bar{x}_{i1} \bar{x}_{i2} \bar{x}_{j1} \Bar{x}_{j2}  \Big]  \tiny{+} \nonumber   \\  \nonumber & c_{\perp} (i,j)\Big[  \sum_{t=i,j}\sum_{n=1,2} x_{tn} \bar{x}_{\bar{t}n} \bar{x}_{t\bar{n}} \bar{x}_{\bar{t}\bar{n}}    \Big]  \tiny{+}  c_{\text{GI}_{\text{wc}}}(i,j) \Big[\sum_{n=1,2}  x_{in}   x_{j\bar{n}}  \bar{x}_{i\bar{n}}   \bar{x}_{jn} \Big] \   \tiny{+} c_{\text{  MISO}}(i,j)  \Big[ \sum_{t=i,j} x_{t1} x_{t2} \bar{x}_{\bar{t}1} \bar{x}_{\bar{t}2} \Big] \tiny{+} \\ &  c_{\text{  MIMO}}(i,j) \Big[x_{i1}   x_{i2}  x_{j1}   x_{j2} \Big] \Big\}         \tiny{+} \sum_{\substack{f_j \in \mathcal{F} \\ f_i = f_j}} q_i^2 \Big\{ c_{\text{ MC}}^{\text{\tiny MBS}}(i,j)[\bar{x}_{i1} \bar{x}_{i2} ]  \tiny{+}c_{\text{\tiny MC}}^{\text{\tiny SBS}}[x_{i1} \bar{x}_{i2}\tiny{+} x_{i2} \bar{x}_{i1}]\tiny{+} c_{\text{GI}_{\text{c}}}(i,j) [x_{i1}x_{i2}]\Big\}.     
\end{align} \vspace{-2.5mm} \hrule 
\end{figure*}

\section{Optimization Problem}\label{sec:optprob}
%\vspace{-0.05in}

\subsection{Cooperative Approach}
 In this section we  present the caching optimization problem for the CA.  In the CA a user can be served by the MBS or any of the SBSs. 
 The expected power cost for serving a request $Q_c(\mathbf{x}_1, \mathbf{x}_2)$  can be written as in (\ref{Qxc}). % where $c_T(i,j)$  denotes the minimum power consumption derived   for the different channels in Section-\ref{sec:Transmissions}.   
 In the following, given a variable $h=\{1,2\}$ we indicate with $\bar{h}$ its complementary value, i.e. if $h=1$ then $\bar{h}=2$ and vice-versa.
Denoting the  capacity of a point to point Gaussian channel with signal-to-noise ratio $x$ as $C(x) = \frac{1}{2}\log_2(1+x)$, the optimization problem for the CA can be formulated as follows   
\begingroup
\allowdisplaybreaks 
 \begin{align}  \label{Qxcooperative} 
&\underset{P^{(1)},P^{(2)},\mathbf{x}_1, \mathbf{x}_2}{\text{minimize}} \displaystyle   Q_\text{c}(\mathbf{x}_1, \mathbf{x}_2),& \\
&\label{eqchace1} \displaystyle {\text{subject to:}}  \sum\limits_{f_i \in \mathcal{F}} x_{in} \leq  M_n,    & n=\{1,2\}, & \\ 
& \label{BC8} \bar{x}_{i1} \cdot \bar{x}_{i2} \cdot \bar{x}_{j1} \cdot \bar{x}_{j2}   \cdot  R_- \leq  C(a_{-0}, P_-  ) 
\end{align} 
\begin{align}
& \label{BC9} \bar{x}_{i1} \cdot \bar{x}_{i2} \cdot \bar{x}_{j1} \cdot \bar{x}_{j2}  \cdot   R_+ \leq C\left(a_{+0}\cdot P_{+}\right), \\  %f_i,f_j \in \mathcal{F}, f_i\neq f_j,     \\ 
& \label{orth10} x_{lw}\cdot \bar{x}_{l\bar{w}} \cdot \bar{x}_{\bar{l}w} \cdot \bar{x}_{\bar{l}\bar{w}}  \cdot R_{\bar{l}} \leq C\left(a_{\bar{w}0}   P_{\bar{l}}^{(\bar{w})}   \right), \\ & \quad \quad \quad \quad \quad \quad \quad   \quad \quad \quad \quad \quad \quad \quad            w=\{1,2\}, l=\{i,j\}, \nonumber \\
& \label{orth17}  x_{lw}\cdot \bar{x}_{l\bar{w}} \cdot \bar{x}_{\bar{l}w} \cdot \bar{x}_{\bar{l}\bar{w}}  \cdot   R_{l} \leq C\left(a_{wk}    P_l^{(w)}  \right), \\ &  \quad \quad \quad \quad \quad \quad \quad   \quad \quad \quad \text{if } l=i \text{ then } k=1 \text{ else } k=2,  \nonumber \\
& \label{HK18} x_{ir}\cdot x_{j\bar{r}} \cdot \bar{x}_{i\bar{r}}\cdot \bar{x}_{jr}    R_i \leq \rho_1(c_{r\bar{r}},c_{\bar{r}r},\tilde{P}_i^{(1)},\tilde{P}_j^{(2)}), \\ 
&\quad \quad \quad \quad \quad \quad \quad   \quad \quad \quad \quad \quad \quad \quad \quad \quad \quad   \quad    
r= \{1,2\},  \nonumber\\  
& x_{ir}\cdot x_{j\bar{r}}\cdot \bar{x}_{i\bar{r}}\cdot \bar{x}_{jr}    R_j \leq \rho_2( c_{r\bar{r}}, c_{\bar{r}r},\tilde{P}_i^{(1)},\tilde{P}_j^{(2)})   \\ 
& x_{ir}\cdot x_{j\bar{r}}\cdot \bar{x}_{i\bar{r}}\cdot \bar{x}_{jr}    (R_i + R_j )\leq \rho_{12}( c_{r\bar{r}}, c_{\bar{r}r},\tilde{P}_i^{(1)},\tilde{P}_j^{(2)})  \\
& x_{ir}\cdot x_{j\bar{r}}\cdot \bar{x}_{i\bar{r}}\cdot \bar{x}_{jr}   ( 2 R_i + R_j) \leq \rho_{10}( c_{r\bar{r}},  c_{\bar{r}r},\tilde{P}_i^{(1)},\tilde{P}_j^{(2)})  \\  
& \label{HK22}x_{ir}\cdot x_{j\bar{r}}\cdot \bar{x}_{i\bar{r}}\cdot \bar{x}_{jr}   ( R_i + 2  R_j) \leq \rho_{20} ( c_{r\bar{r}}, c_{\bar{r}r},\tilde{P}_i^{(1)},\tilde{P}_j^{(2)}) \\
& \label{MIMO28}  x_{i1} \cdot x_{i2} \cdot x_{j1} \cdot  x_{j2} \cdot R_i \leq C(\pi_m, \mathbf{H}_1, \mathbf{H}_2, \mathbf{S}_1, \mathbf{S}_2),  \\
& \label{MIMO29}  x_{i1} \cdot  x_{i2} \cdot x_{j1} \cdot  x_{j2}  \cdot R_j \leq    C(\pi_m, \mathbf{H}_1, \mathbf{H}_2, \mathbf{S}_1, \mathbf{S}_2 ) \\
& \label{BCSBS30}x_{iz}\cdot x_{jz}\cdot \bar{x}_{i\bar{z}} \cdot \bar{x}_{j\bar{z}} \cdot  R_- \leq  C\left(a_{-z} P_{-}/(1+a_{-z} P_{+})\right), \\ &  \quad \quad \quad \quad \quad \quad \quad   \quad \quad \quad \quad \quad \quad \quad \quad \quad \quad     \quad z=\{1,2\}, \nonumber \\ 
& \label{BCSBS31}x_{iz}\cdot x_{jz}\cdot \bar{x}_{i\bar{z}} \cdot \bar{x}_{j\bar{z}} \cdot  R_+ \leq C\left(a_{+z} P_{+}\right),   \\  
& \label{MISO34}x_{i1}\cdot x_{i2}\cdot \bar{x}_{j1} \cdot \bar{x}_{j2}  \cdot  R_i \leq C\left( (a_{11}+a_{12}) \cdot P_i^{(1)} \right), \\
& \label{MISO35}x_{i1}\cdot x_{i2}\cdot \bar{x}_{j1} \cdot \bar{x}_{j2}  \cdot  R_j \leq C(a_{20}\cdot P_j^{(2)}),   \\
& \label{MISO36}x_{j1}\cdot x_{j2}\cdot \bar{x}_{i1} \cdot \bar{x}_{i2} \cdot  R_i \leq C(a_{10} \cdot P_i^{(1)}),   \\
&  \label{MISO37}x_{j1}\cdot x_{j2}\cdot \bar{x}_{i1} \cdot \bar{x}_{i2} \cdot  R_j \leq C\left ( (a_{22}+a_{21}) \cdot P_j^{(2)} \right),  \\ 
& \label{MCL38}x_{ip}\cdot \bar{x}_{i\bar{p}} \cdot R_i \leq C(a_{pp}  \cdot P_i^{(p)}), \quad \quad \quad \quad \quad  p=\{1,2\}, \\
& \label{MCL39}x_{ip}\cdot  \bar{x}_{i\bar{p}}  \cdot R_i \leq C(a_{\bar{p}p} \cdot P_j^{(\bar{p})}),   \\
& \label{MCMBS44}\bar{x}_{i1} \cdot \bar{x}_{i2}\cdot \bar{x}_{j1} \cdot\bar{x}_{j2} \cdot R_i \leq C\left(a_{10} \cdot P_i^{(1)}\right), \quad \quad  i = j, \\ 
& \label{MCMBS45}\bar{x}_{i1}  \cdot \bar{x}_{i2} \cdot \bar{x}_{j1}\cdot \bar{x}_{j2} \cdot R_j \leq C\left(a_{20} \cdot P_j^{(2)}\right), \quad        \quad i = j,   \\
& \label{IC42}x_{i1}\cdot x_{i2} \cdot R_i  \leq C\Big(\Big(\sqrt{\tilde{P}_i^{(1)}}+\sqrt{c_{21}\cdot \tilde{P}_j^{(2)}}\Big)^2 \Big),    i = j,  \\
& \label{IC43}x_{i1}\cdot x_{i2} \cdot R_j  \leq C\Big(\Big(\small{\sqrt{\tilde{P}_j^{(2)}}}+\small{\sqrt{c_{12}\cdot \tilde{P}_i^{(1)}}}\Big)^2 \Big),      i = j,  \\
&\label{P46}\displaystyle  P_i^{(1)} \geq 0, P_j^{(2)} \geq 0,& \\ 
 & P_-  \geq 0, P_+  \geq 0,   \\
% &\label{Pmax36} P_i^{(1)} \leq P_{\text{MAX}}, P_j^{(2)} \leq P_{\text{MAX}}, P_+ \leq P_{\text{MAX}}, P_- \leq P_{\text{MAX}, 
&\label{x48}\displaystyle  x_{in}\in \{0,1\}, x_{jn}\in \{0,1\}, \textcolor{white}{letr} & &\\
& \label{fifj}f_i, f_j \in \mathcal{F}. & 
\end{align} 
\endgroup
\normalsize
Inequality (\ref{eqchace1}) ensures that the total amount of data stored in a cache does not exceed its size, (\ref{BC8})-(\ref{BC9}) denote the
capacity region of the Gaussian broadcast channel of the MBS, while (\ref{orth10})-(\ref{orth17}) represent the conditions for the case of orthogonal channels,
(\ref{HK18})-(\ref{HK22}) are the Han-Kobayashi conditions and define the best known achievable rate of the GI$_{wc}$ (the explicit expressions for
such conditions are given in the Appendix II), (\ref{MIMO28})-(\ref{MIMO29}) define the capacity region for the MIMO broadcast channel, (\ref{BCSBS30})-(\ref{BCSBS31}) %and (\ref{BCSBS32})-(\ref{BCSBS33}) 
   denote the capacity region for the broadcast channel of SBS$_n$, (\ref{MISO34})-(\ref{MISO35}) and (\ref{MISO36})-(\ref{MISO37}) define the capacity region of the orthogonal channel when the SBSs implement a MISO channel to serve $u_1$ ($u_2$) while $u_2$ ($u_1$) is served by the MBS, (\ref{MCL38})-(\ref{MCL39}) % and (\ref{MCL40})-(\ref{MCL41})
    denote the capacity region of the multicast channel of SBS$_n$, (\ref{IC42})-(\ref{IC43}) define an achievable rate region of the GIc when both SBSs
have to transmit the same file, (\ref{MCMBS44})-(\ref{MCMBS45}) define the capacity region of the multicast channel of the MBS, (\ref{P46}) is imposed
to guarantee the non negativity of the transmit power, while  (\ref{x48}) represents the caching choice and accounts for the discrete
nature of the optimization variable.  
Note that the power variables
in constraints (\ref{HK18})-(\ref{HK22}) and (\ref{IC42})-(\ref{IC43}) refer to the normalized channel model ($\tilde{P}_i^{(1)}$, $\tilde{P}_j^{(2)}$). Since we are interested in evaluating
the physical power consumption, the power values in the
calculation of the cost have to be scaled as $P_i^{(1)}=\tilde{P}_i^{(1)}/a_{11} $ and $P_j^{(2)}=\tilde{P}_j^{(2)}/a_{22} $.
As a side remark, note that there is a significant difference between the MISO and the GI$_c$ channel, despite the fact that in both cases the same file is transmitted by both SBSs. In fact, in the GI$_c$ channel SBSs  transmit to both users and powers are adjusted such that both users achieve the QoS required while in the MISO channel the SBSs transmit only to one user.    
\vspace{-.05in}
\subsection{Non-Cooperative Approach}   
\vspace{-.01in}
 In the NCA each user is served either by  the SBS to which it is associated or by the MBS. This approach considers for all transmissions the interference of the other SBS  as noise. The MBS transmits to the user  when the file requested by $u_n$  is not present in the cache of SBS$_n$  or when the channel conditions do  not allow to have a reliable transmission.
The expected power cost  in the NCA $Q_{nc}(\mathbf{x}_1, \mathbf{x}_2)$ can be written as in (\ref{QnonC}). 
 The optimization problem to be  solved is the following 
 \begin{figure*}[t]\normalsize
\centering  
 \begin{align}  \label{QnonC} 
& Q_{\text{nc}}(\mathbf{x}_1, \mathbf{x}_2) \tiny{=}  
  \sum_{f_i \in \mathcal{F}} \sum_{\substack{f_j \in \mathcal{F}\nonumber   \\ f_i\neq f_j}} q_{i}   q_{j} \Big\{  c_{\text{GIN}}(i,j)  \big[ \mathbbm{1_{\alpha}}   x_{i1}x_{j2} \big]    \tiny{+}   c_{\perp}(i,j)   \big[ x_{i1} \bar{x}_{j2} + \bar{x}_{i1}  x_{j2} \big]  \tiny{+}   c_{\text{ BC}}^{\text{\tiny MBS}}(i,j)   \big[\mathbbm{\bar{1}_{\alpha}}   x_{i1}x_{j2}        \\ 
&  \tiny{+} \bar{x}_{i1} \bar{x}_{j2}  \big] \Big\}       \tiny{+}  \sum_{\substack{f_j \in \mathcal{F} \\ f_i = f_j}} q_{i}^2 \Big\{ c_{\text{GIN}}(i,j)     \big[\mathbbm{1_{\alpha} }   x_{i1}    x_{i2}\big] \tiny{+}  c_{\text{ MC}}^{\text{\tiny MBS}}(i,j) [\bar{x}_{i1}   \bar{x}_{i2}     \tiny{+}    \mathbbm{\bar{1}_{\alpha}}   x_{i1}   x_{i2}\big]    \tiny{+}    c_{\perp}(i,j) \big[   \bar{x}_{i2}    x_{i1}    \tiny{+}   \bar{x}_{i1}   x_{i2}\big] \Big\}.       
\end{align}   
  
 \hrule \ 
\end{figure*} %\vspace{-0.3in}

\begingroup 
\allowdisplaybreaks
 \begin{align} 
& \underset{P^{(1)},P^{(2)},\mathbf{x}_1, \mathbf{x}_2}{\text{minimize}} \displaystyle   Q_{\text{nc}}(\mathbf{x}_1, \mathbf{x}_2), \\
&\text{subject to: (\ref{eqchace1})-(\ref{BC9}), (\ref{MCMBS44})-(\ref{fifj})} \nonumber  \\
 &\label{ncBC56} x_{kg} \cdot \bar{x}_{\bar{k}g}\cdot \bar{x}_{\bar{k}\bar{g}}\cdot \bar{x}_{k\bar{g}} \cdot   R_+ \leq C\left(a_{+0} P_{+}\right), \\ 
 & \quad \quad \quad \quad \quad \quad \quad       g=\{1,2\}  {\text{  if } g =  1 \text{ then } k=j \text{ else } k=i}    \nonumber   \\ 
& \label{ncBC55} x_{kg} \cdot \bar{x}_{\bar{k}g}\cdot \bar{x}_{\bar{k}\bar{g}} \cdot \bar{x}_{k\bar{g}}  \cdot R_- \leq  C\left(a_{-0} P_{-}/(1+a_{-0}  P_{+})\right),  \\
& \label{ncBC53}x_{i2}\cdot x_{j1}\cdot  \bar{x}_{i1}\cdot \bar{x}_{j2} \cdot   R_- \leq  C\left(a_{-0} P_{-}/(1+a_{-0} P_{+})\right), \\
& \label{ncBC54}x_{i2}\cdot x_{j1}\cdot \cdot \bar{x}_{i1}\cdot \bar{x}_{j2} \cdot  R_+ \leq C\left(a_{+0} P_{+}\right), \\ &
 \label{orth10n} x_{lw} \cdot \bar{x}_{l\bar{w}}\cdot \bar{x}_{\bar{l}w}\cdot \bar{x}_{\bar{l}\bar{w}} \cdot  R_{\bar{l}} \leq C\left(a_{\bar{w}0} \cdot  P_{\bar{l}}^{(\bar{w})}   \right) ,   \\ &   \quad \quad \quad \quad \quad \quad \quad                 w=\{1,2\} \normalsize{\text{  if } w = 1 \text{ then } l=i \text{ else } l = j}         \nonumber \\  &
 \label{orth17n}  x_{lw}\cdot \bar{x}_{l\bar{w}}\cdot \bar{x}_{\bar{l}w} \cdot \bar{x}_{\bar{l}\bar{w}} \cdot R_{l} \leq C\left(a_{ww}  \cdot P_l^{(w)}  \right),  \\&
 \label{orth57}  x_{tm} \cdot x_{\bar{t}m}\cdot \bar{x}_{t\bar{m}}\cdot \bar{x}_{\bar{t}\bar{m}}  \cdot R_t \leq  C(a_{tt}  \cdot  P_t^{(m)}), \\ & \quad \quad \quad \quad \quad \quad \quad \quad \quad \quad \quad  \quad \quad \quad \quad   t=\{i,j\}, m=\{1,2\},  \nonumber  \\  &
 \label{orth58} x_{tm} \cdot x_{\bar{t}m}\cdot  \cdot \bar{x}_{t\bar{m}}\cdot \bar{x}_{\bar{t}\bar{m}}  \cdot R_{\bar{t}}\leq   C(a_{\bar{t}0} \cdot P_{\bar{t}}^{(\bar{m})}),  \\ &
 \label{orth61}x_{tm}\cdot x_{t\bar{m}}\cdot \bar{x}_{\bar{t}m}\cdot \bar{x}_{\bar{t}\bar{m}}  \cdot  R_t \leq C(a_{mm}  \cdot  P_t^{(m)}),  \\ & 
 \label{orth62}x_{tm}\cdot x_{t\bar{m}}\cdot \cdot \bar{x}_{\bar{t}m}\cdot \bar{x}_{\bar{t}\bar{m}}   \cdot  R_{\bar{t} }\leq C(a_{\bar{m}0}\cdot P_{\bar{t}}^{(\bar{m})}),  \\ &
 \label{GIN65}\mathbbm{1_{\alpha}}\cdot  x_{i1} \cdot x_{j2}\cdot \bar{x}_{i2}\cdot \bar{x}_{j1}  \cdot R_i \leq  C\left(a_{11}     P_i^{(1)}/(1 + a_{12}  P_j^{(2)})\right),  \\ &
 \label{GIN66} \mathbbm{1_{\alpha}} \cdot x_{i1} \cdot x_{j2}\cdot \bar{x}_{i2}\cdot \bar{x}_{j1}  \cdot R_j \leq C\left(a_{22}    P_j^{(2)}/(1 + a_{21}  P_i^{(1)})\right), \\&
  \label{GIN67}\mathbbm{1_{\alpha}} \cdot x_{i1} \cdot x_{i2} \cdot x_{j1} \cdot  x_{j2} \cdot R_i \leq  C\left(a_{11}    P_i^{(1)}/(1 + a_{12}  P_j^{(2)})\right), \\&
 \label{GIN68} \mathbbm{1_{\alpha}} \cdot  x_{i1} \cdot  x_{i2} \cdot x_{j1} \cdot  x_{j2}  \cdot R_j \leq C\left(a_{22} P_j^{(2)}/(1 + a_{21}  P_i^{(1)})\right),   \\&
 \label{GIN75mc} {\bar{1}_{\alpha}} \cdot x_{i1} \cdot x_{i2} \cdot x_{j1} \cdot  x_{j2} \cdot R_i \leq  C\left(a_{10} P_i^{(2)}\right), \quad i=j, \\ &
 \label{GIN76mc}\mathbbm{\bar{1}_{\alpha}} \cdot  x_{i1} \cdot  x_{i2} \cdot x_{j1} \cdot  x_{j2}  \cdot R_j \leq  C\left(a_{20} P_j^{(2)}\right), \quad i=j,  \\&
 \label{GIN73mc}{\bar{1}_{\alpha}}\cdot x_{i1}\cdot x_{j2} \cdot \bar{x}_{i2}\cdot \bar{x}_{j1} \cdot  R_i \leq  C\left(a_{10} P_i^{(1)}\right),\quad i=j,   \\ &
 \label{GIN74mc} {\bar{1}_{\alpha}}\cdot x_{i1}\cdot x_{j2} \cdot\bar{x}_{i2} \cdot \bar{x}_{j1} \cdot R_j \leq C\left(a_{20} P_j^{(2)}\right), \quad   i=j,    \\ &
  \label{GIN71bc} \mathbbm{\bar{1}_{\alpha}} \cdot  x_{i1} \cdot  x_{i2} \cdot  x_{j1}\cdot    x_{j2}  \cdot R_- \leq  C\left(a_{-0} P_{-}/(1+a_{-0} P_{+})\right)  \\ &
 \label{GIN72bc}\mathbbm{\bar{1}_{\alpha}} \cdot  x_{i1} \cdot  x_{i2} \cdot x_{j1} \cdot  x_{j2}  \cdot R_+ \leq C\left(a_{+0} P_{+}\right)       \\ &
 \label{GIN69bc}  \mathbbm{\bar{1}_{\alpha}}  \cdot x_{i1}   \cdot  x_{j2} \cdot \bar{x}_{i2} \cdot \bar{x}_{j1} \cdot   R_- \leq  C\left( a_{-0} P_{-}/(1+a_{-0} P_{+})\right),           \\ & 
 \label{GIN70bc} \mathbbm{\bar{1}_{\alpha}} \cdot  x_{i1}\cdot x_{j2} \cdot \bar{x}_{i2}  \cdot \bar{x}_{j1} \cdot R_+ \leq C\left(a_{+0} P_{+}\right).
  \end{align} \vspace{-0.05in}
\endgroup   
    
Conditions (\ref{eqchace1})-(\ref{BC9}) and (\ref{MCMBS44})-(\ref{fifj}) are in common with the CA. Conditions (\ref{ncBC56})-(\ref{ncBC54})  denote the capacity region of the Gaussian broadcast channel, (\ref{orth10n})-(\ref{orth62})
 define the capacity region of the orthogonal channels, while conditions (\ref{GIN65})-(\ref{GIN68}) represent  the constraints of the GIN channel. Constraints 
 (\ref{GIN75mc})-(\ref{GIN74mc}) and  (\ref{GIN71bc})-(\ref{GIN70bc}) denote the capacity region of the Gaussian multicast channel considered  when the SBSs are not capable to deliver the files at the requested rates.
 \vspace{-0.1in}
\subsection{Implementation}\label{sec:imple} 
 The optimal cache allocation can be derived by  minimizing the cost function in (\ref{Qxc}) independently along the power dimension and the cache allocation dimension. First, the minimum average power for a given cache allocation is  calculated. Let {us recall} that $N$ is the number of files. For each of the $N^2$ possible requests the power  has to be minimized. Each request, according to whether the file is present or not in a cache, requires to solve one of the power minimization subproblems presented in Section III.    
  The optimal cache allocation is obtained by exhaustive search within the set of possible allocations.
  The  complexity  for minimizing the average power for a given cache allocation is $O(N^2)$ while the cost for finding the optimal cache  ${\text{allocation is } O(N!)}$.    

   \begin{algorithm}[t]
 \caption{\small $[\textbf{x}_1 \quad \textbf{x}_2] =$ low\_complexity\_allocation($\textbf{R}, \textbf{q}$)}\label{algo}
  \begin{algorithmic}[1]
 \small
  \State $\textbf{x}_1 = \textbf{0}$ and $\textbf{x}_2  = \textbf{0} \quad  \quad $  \Comment {$\textbf{0}$ is an $N$-dimensional all-zero vector}
   \State $W = \left\{ w_1, \dots,  w_k, \dots, w_N \right\}$   \Comment $w_k = 2^{2 R_k} \cdot q_k$ is the  $k^{th}$ element of the set
   \For{$l = 1, \dots , M$}
  \State $w_m$ = $\underset{l} \arg\max\{w_l\}$
   \State $x_{m1}=1$ 
   \State   $W = W \setminus \{w_m\}$  
    \State $w_m$ = $\underset{l} \arg\max\{w_l\}$
   \State $x_{m2}=1$  
      \State   $W = W \setminus \{w_m\}$
  \EndFor
 \State \textbf{return} $[\textbf{x}_1 \quad \textbf{x}_2]$ 
  \end{algorithmic}
 \end{algorithm}

Although such algorithm is useful to evaluate the performance limit of the considered setup, its complexity makes it unsuited for a practical application when the number of files is large. Therefore, we propose in the following a suboptimal algorithm with   reduced   complexity.

Let us consider equation (\ref{Qxc}). The cost of each channel depends in a non-linear way on the channel coefficients $a_{nm}$ and on the rate  of both requested files $R_i, R_j$. It also depends linearly on the probabilities of the files requests $q_i, q_j$. Let us now consider genie-aided receivers having access to the message destined to the other user. In this case, the interference can be entirely removed and each user sees an interference-free channel. In such channel, the cost of transmitting  $f_i$ is equal to its popularity ranking $q_i$ times a term proportional to $2^{R_i}$. 
Since we consider a setup in which the SBS-user channel is on average better than that between MBS-user one, having files with higher cost transmitted by SBSs rather than the MBS minimizes the overall power expenditure. 

Let us define the utility function {$w_k(R_k, q_j) =  2^{2 R_k} \cdot q_k$} associated to $f_k$.
The proposed algorithm, described in Algorithm \ref{algo}, caches the files with highest $w(R, q)$. The  complexity of the algorithm grows linearly ${\text{with } N}$. In section V, we compare this algorithm with the optimal one and with two benchmarks. 
\section{Numerical Results}\label{sec:num_res}%
In this section we compare CA and NCA in two different scenarios, namely: (i) static scenario and  (ii) mobile transmitters scenario. 
In the static scenario no fading is present, i.e., all channel coefficients are fixed. In the mobile transmitters scenario the transmitter-user links experience fading while the master node-user links are AWGN channels. This models a heterogeneous satellite network in which  two LEO satellites act as transmitters while a GEO satellite acts as master node. Due to the LEOs orbital motion and the presence of obstacles such as tall buildings, fading is present in the LEO-user channels\footnote{In practice the equivalent of an SBS (MBS) would be a LEO (GEO) satellite together with the respective ground station, where the cache would be physically located. This is because on-board storage in satellites is not common practice.} while the link to the GEO, assumed to be line of sight, does not change significantly over time, which is the case in clear-sky conditions. This setup also models a hybrid network with a GEO satellite acting as master node while hovering UAVs play the role of transmitters. Also in this case, the UAV-user links are affected by fading due to UAVs motion \cite{UAVchMod}.
\begin{center}
\begin{table}[htp] \centering
\caption{Rate and popularity rankings for each file in the direct probability and inverse probability setups.}\label{tab:table}
\begin{tabular}{lllr}
\hline
$\mathcal{F}$ & $R_i$ [bits/sec/Hz]       & $q_i$ \emph{direct} &  $q_i$ \emph{inverse}  \\
\hline
$f_1 $ & 1.2      & 0.45   & 0.05      \\
$f_2 $ &1.0      & 0.20   & 0.15       \\
$f_3 $ &0.5      & 0.15     & 0.15      \\
$f_4 $ &0.4      & 0.15     & 0.20      \\
$f_5 $ &0.2      & 0.05      & 0.45      \\
\hline 
\end{tabular}
\end{table}
\end{center}
\vspace{-0.4in} 
 We model the channel coefficients in the links affected by fading as Lognormal random variables, i.e., given a channel coefficient $a$, we have $\log(a)\sim\mathcal{N}\left(\mu, \sigma^2\right)$, $\log(.)$ being the Napierian logarithm. The parameters $\mu$ and $\sigma$ have been chosen so that in the three scenarios  we have $\mathbb{E}\{a_{11} \} = \mathbb{E}\{a_{22} \} = 1$ and   $ \mathbb{E}\{a_{10}\}=\mathbb{E}\{a_{20}\}=$ $0.01$, $\mathbb{E}\{.\}$ being the expectation operator. 
We plot our results against the mean value of the interfering link coefficients as presented in the  GIc equivalent model ($c_{12}$, $c_{21}$), where direct channel coefficients have unit mean. The parameters for such coefficients are chosen so that $\mathbb{E}\{c_{12}\}=\mathbb{E}\{c_{21}\}$ take values in $[0,1]$. This allows to compare different scenarios against a single parameter. In order to do a fair comparison, this is done for both the non-cooperative as well as the cooperative scheme. 
%\vspace{-0.3in}

We consider the system depicted in Fig. \ref{fig:mod_nc} for the NCA and Fig.~\ref{fig:mod_coop} for the CA where each SBS  has a memory of size $M=2$ and  the library has cardinality $N=5$.  
Two relevant setups are considered. In the first one, files with higher rate have a higher popularity ranking while in the second one files with higher rate have a lower popularity ranking. We refer to these setups as \emph{direct probability} and \emph{inverse probability}, respectively. The corresponding values for $R_i$ and $q_i$ are shown in Table \ref{tab:table}. %%%%%%%%%
 We plot for each scenario  the minimum power cost $Q(\mathbf{x}_1, \mathbf{x}_2)$ of the most significant cache allocations in the CA (solid curves) and in the NCA (dashed curves) versus the mean value of the interference coefficients, assumed to be the same for both users\footnote{Note that, while in the static case this is equivalent to having a symmetric channel matrix, this is not necessarily the case in the mobile transmitter case due to the independence of fading.}.  Note that, by the symmetry of the problem, the results do not change if the content of the cache of SBS$_1$ and that of SBS$_2$ are switched, i.e., the performance of the solution $\mathcal{M}_1=\{f_i,f_j\}$,  $\mathcal{M}_2=\{f_k,f_l\}$  and $\mathcal{M}_1=\{f_k,f_l\}$, $\mathcal{M}_2=\{f_i,f_j\}$ coincide.
\begin{figure}[t]
\centering
\includegraphics[width=0.5\textwidth]{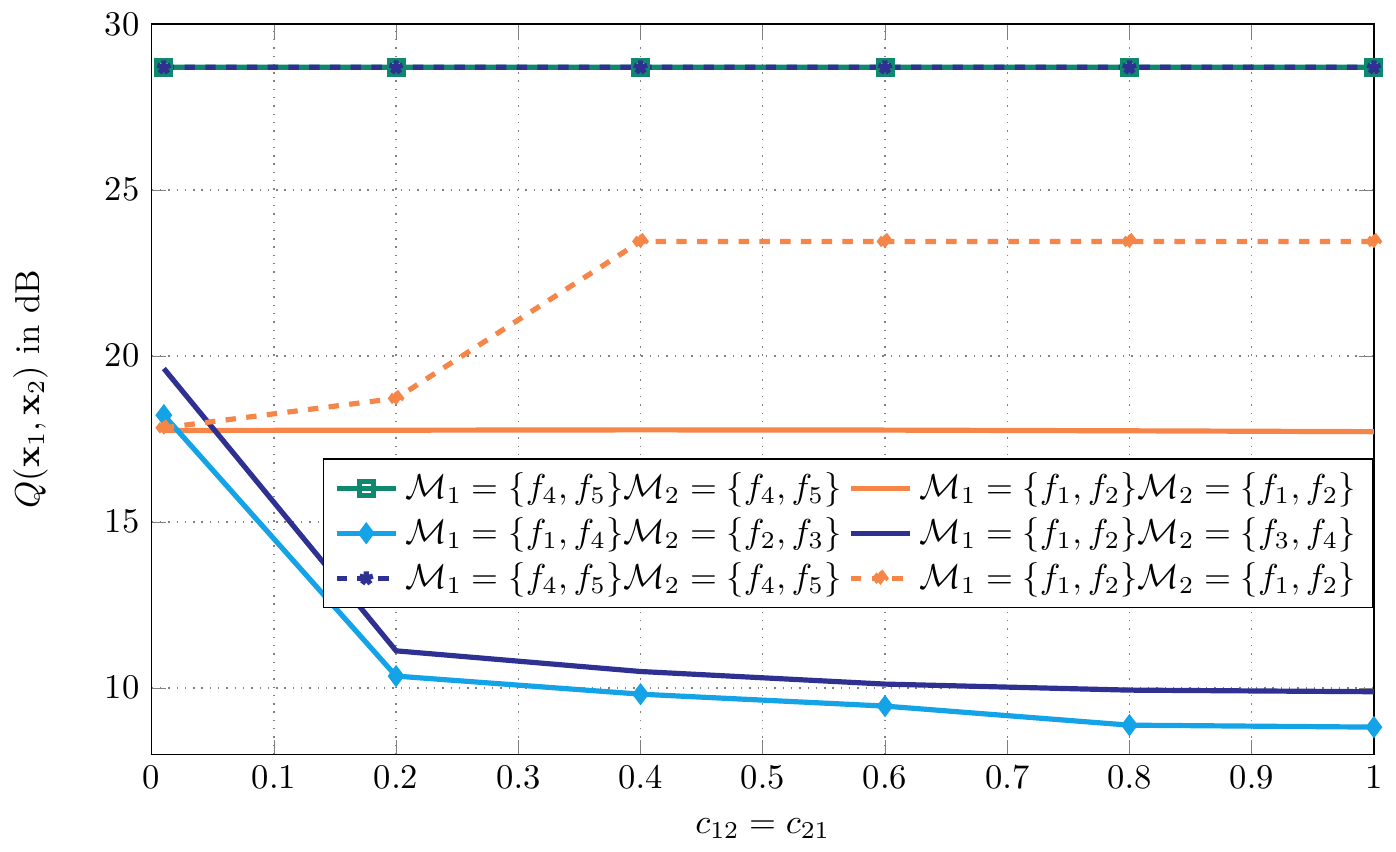}
\caption{Minimum average power cost in dB in the static case plotted against the interference coefficient with direct file probabilities. The result for the CA is represented with solid curves while  dashed curves are used for the the NCA.
The optimization is done for each average interference coefficient. In the simulations we set $a_{10} = a_{20}\ = 0.01$ and $ a_{11} =  a_{22}\ = 1$.}
\label{fig:AGWNdirect}
\end{figure}

 In Fig. \ref{fig:AGWNdirect}  the most significant cache allocations for the static scenario with direct probabilities are plotted.
When  no cooperation is in place the optimal caching solution consists in each  SBS storing the files with highest probabilities/transmission rate ($\mathcal{M}_1= \mathcal{M}_2=\{f_1,f_2\}$) which is  indicated    with the orange dashed curve. In this case, when mutual interference is low the MBS  only transmits files with low probability and low transmission rate. On the contrary, when the level of interference increases the  SBSs are not able to serve their own users even if they have the requested files and transmissions are delegated to the MBS. Thus, the usage of the MBS becomes more frequent and more power is consumed due to the high rate of the files and lower channel gain.  This explains the step of roughly 4 dB from $c_{12}=c_{21}=0.2$ to  $c_{12}=c_{21}=0.4$. For $c_{12}=c_{21}$ larger than or equal to $0.4$ the curve flattens out because, from that point on, the mutual interference is so strong that the required rates cannot be achieved by the SBSs and the MBS takes over all transmissions.
\begin{figure}[t!]
\centering
\includegraphics[width=0.5\textwidth]{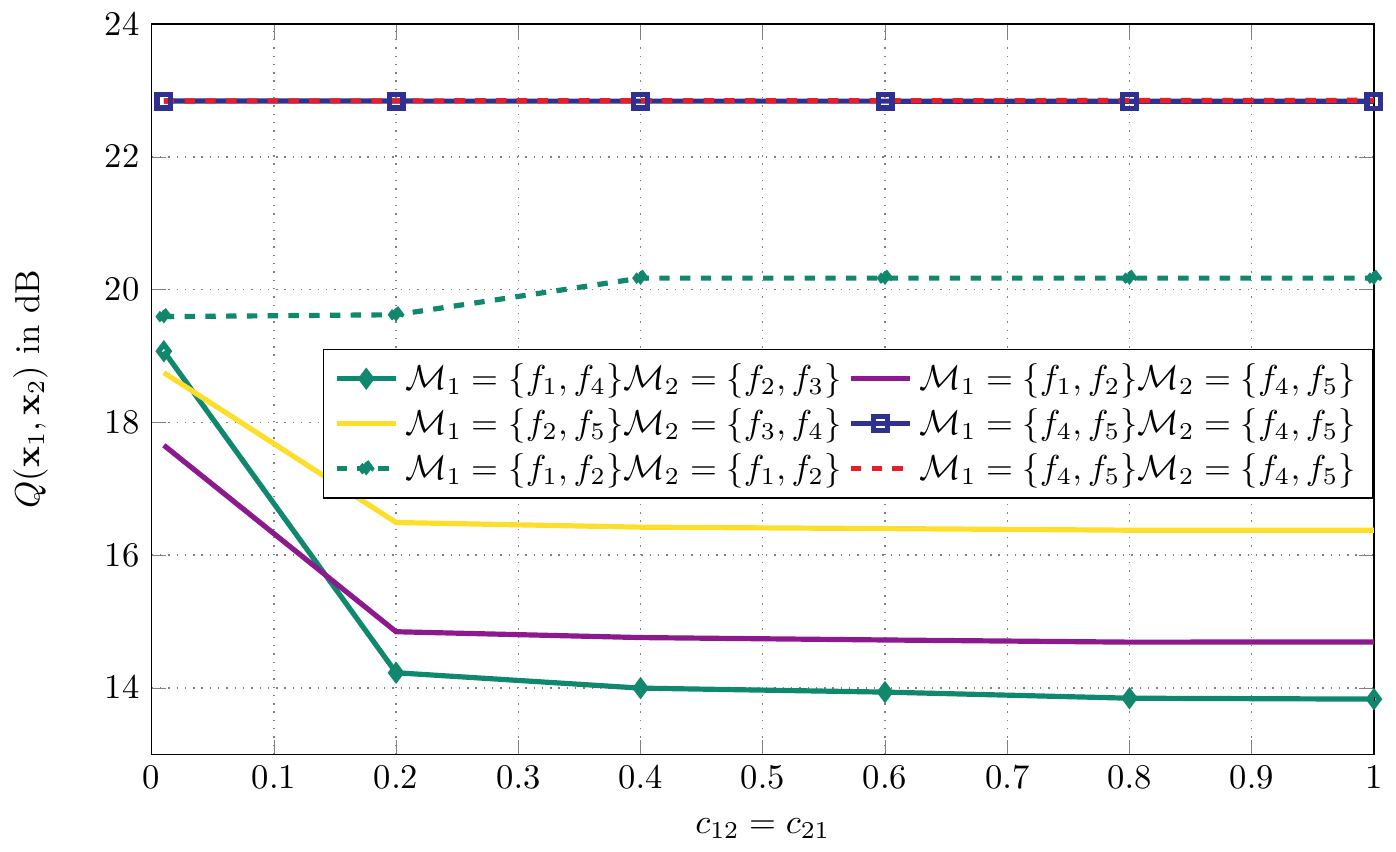}
\caption{Minimum average power cost in dB in the static scenario plotted against the interference coefficient with inverse probabilities.
The optimization is done for each average interference coefficient. In the simulations we set $a_{10} = a_{20}\ = 0.01$ and $ a_{11} =  a_{22}\ = 1$.}
\label{fig:AGWNinverse}
\end{figure}
In the CA the best performance is obtained when one of the SBS stores the files with highest probability/transmission rate and the two caches store  different files. The result is shown in the light blue curve with diamonds. 
 A slight increase in power (from $0.5$ dB up to $1$ dB) is observed when the caches store  different files but the two most probable files are in the same cache ($\mathcal{M}_1=\{f_1, f_2\}$,  $\mathcal{M}_2=\{f_3,f_4\}$). In both CA and NCA approaches, the most power consuming solution coincides and corresponds to each SBS storing the two files with lower probabilities/rates ($\mathcal{M}_1= \mathcal{M}_2=\{f_4,f_5\}$). The flatness of the curves in both the CA and the NCA are due to the fact that the  power consumed by the MBS has a strong weight in the overall mean, since transmissions at high data rate from MBS are the most probable.

 Let us now consider the low interference regime. In the CA,  all the caching schemes  at the lowest interference level consume at least $5$ dB more with respect to the other levels. This is because, when a single SBS   serves both users, it applies either the broadcast or the multicast channel and such channels are penalized when one of the two channel coefficients are relatively small. Later on in this section we show that, despite the higher power consumed in some cases, cooperation has an advantage in terms of MBS resources sparing.
 \begin{figure}[t]
\centering
\includegraphics[width=0.5\textwidth]{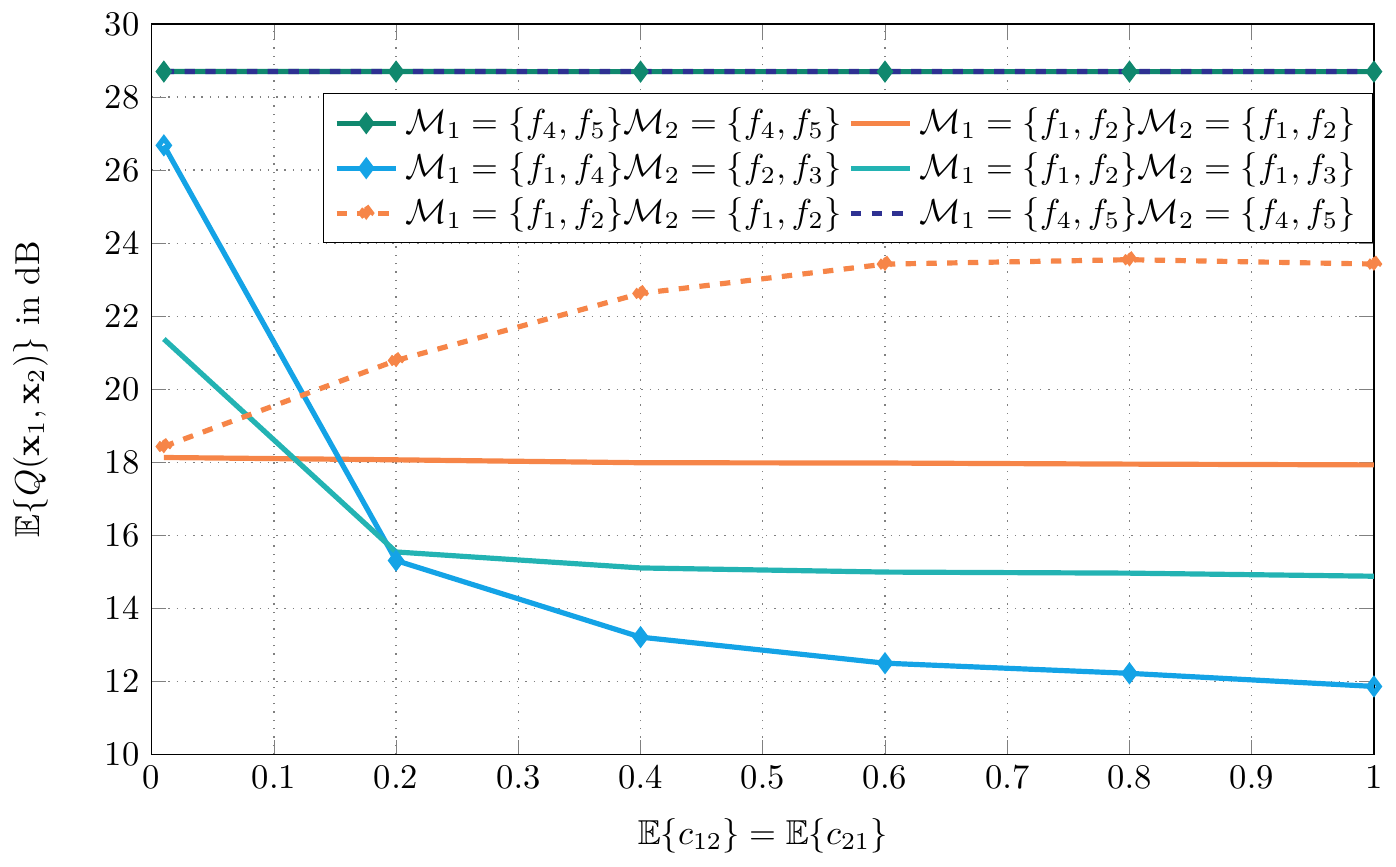}
\caption{Average power cost in dB versus interference link coefficients for different memory allocations in the mobile transmitters setup in a fading environment with direct probabilities. In the simulations we set $a_{10}= a_{20}\ = 0.01$ and  $\mathbb{E}\{{a_{11}\}=\mathbb{E}\{a_{22}\} = 1}$.}
 \label{fig:FIXdirect}
\end{figure}

In Fig.~\ref{fig:AGWNinverse} the power cost versus interference level in the static scenario and for inverse probabilities  (i.e. files with higher $R_i$ have lower $q_i$) is  plotted. The plot shows  that for both approaches the most power-efficient cache allocation (solid and dashed green curves) is not the one including the file with the highest popularity  but rather the one with the highest rate.
 The reason for this is that the weight of MBS transmissions on the overall power budget is higher  with respect to the SBSs' due to the lower channel gains. For the considered rates and popularity ranking it is better to have the MBS transmitting low rate messages more often rather than transmitting   higher rate messages with lower popularity.
   In the cooperative case, if the file with highest rate is not  cached  there is a loss larger than or equal to $2.5$ dB with respect to the best curve. This holds also for the allocation including the most popular file (yellow curve), which confirms that it is better to store the file with highest rate rather than the one with highest popularity.
 As a matter of facts, the worst performance in the both the CA and the NCA case is obtained when each SBS stores the two files with largest probabilities/lowest data rate. 
 
 We examine now the  mobile transmitters scenario (fading on the the SBS-user link, no fading on the the MBS-user link). Note that, due to the difference between two Lognormal random variables at the denominator of equation (\ref{PjIC}), the average transmission power in the GIN channel is not finite\cite{sumLN}. Comparing the CA and the NCA in such conditions always leads to an infinite power gain of the former with respect to the latter as long as the probability that the two SBSs transmit together is non-zero. In order to get some interesting insight in the presence of fading, in the simulations of the mobile transmitters  scenario  we introduced a constraint on the maximum transmit power. Such constraint is chosen so that the probability that the SBSs cannot transmit due to it is around $10^{-5}$ in the worst case scenario (i.e., GIN with maximum rates and $\mathbb{E}\{c_{12}\}=\mathbb{E}\{c_{21}\}=0.4$\footnote{Note that $\mathbb{E}\{c_{12}\}=\mathbb{E}\{c_{21}\}=1$ would lead to a higher outage probability for the SBSs in the NCA, but the outage in such case is due almost exclusively to the lack of a feasible solution to equation (\ref{PjIC}) rather than to the constraint on the maximum power.}).  

 \begin{figure}[t]
\centering
\includegraphics[width=0.5\textwidth]{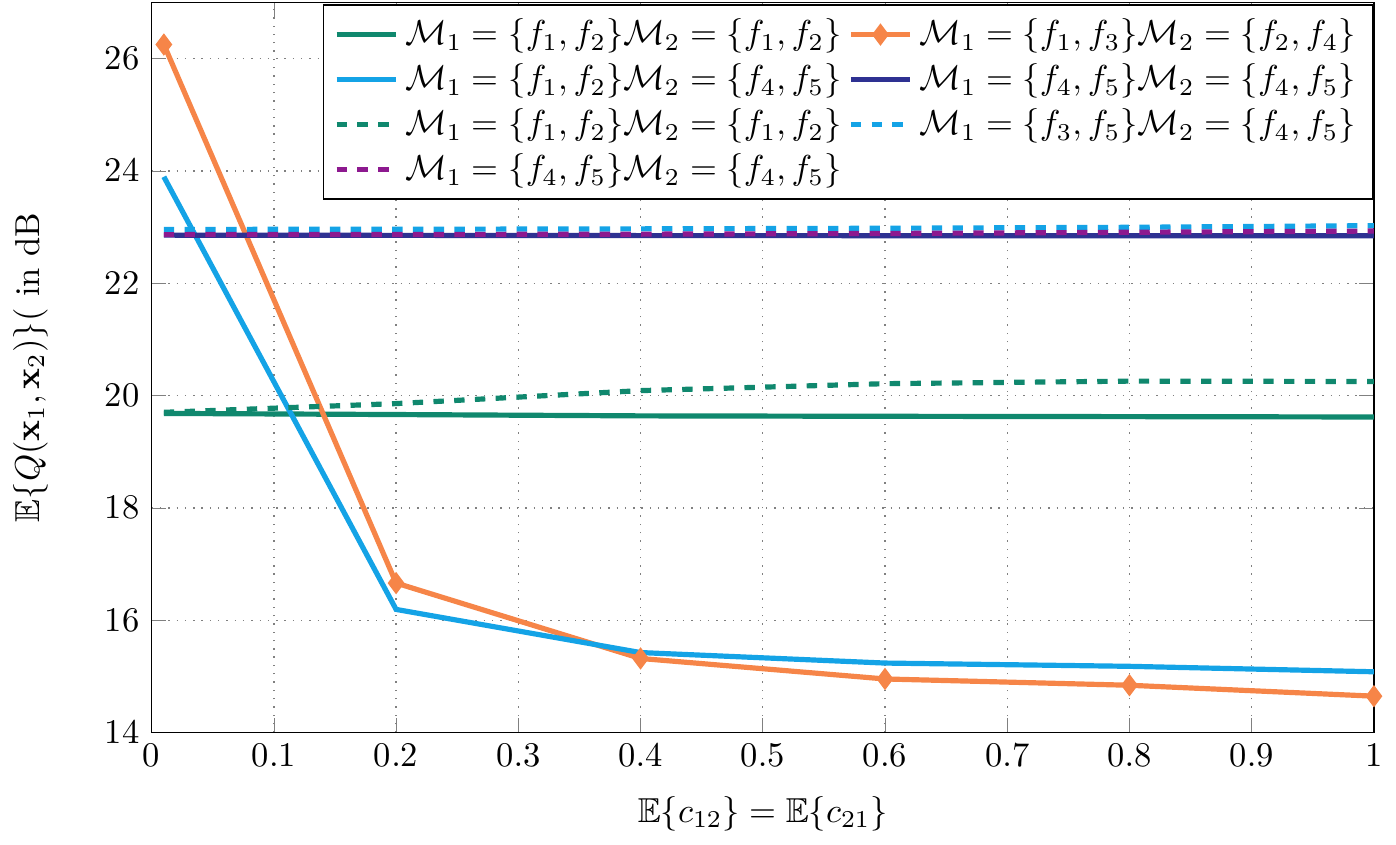}
\caption{Avarege power cost in dB versus interference link coefficients for different memory allocations for the mobile transmitters setup  in a fading environment with inverse probabilities. In the simulations we set $a_{10}= a_{20}\ = 0.01$ and  ${\mathbb{E}\{a_{11}\}
=\mathbb{E}\{a_{22}\} = 1}$.}
\label{fig:FIXinverse}
\end{figure}
 
 In Fig. \ref{fig:FIXdirect}  the minimum   cost  $ \mathbb{E}\{Q(\mathbf{x}_1, \mathbf{x}_2)\}$   for different average interference levels in the mobile transmitters case for direct probabilities is plotted. Let us first consider the CA. Similarly to the static scenario, when  interference is not close to zero  the best performance is obtained when  each SBS stores different files and the most probable ones are in $\mathcal{M}_1 \cup\mathcal{M}_2$. The power consumption is constant ($18$ dB) for all levels of interference when the caches of the SBSs are equal and each SBS  stores the files with highest rate/popularity. 
Note that such caching solution is  the best for very low levels of interference. In fact, when both SBSs cache  the same files, broadcast and multicast transmissions, which are the most expensive ones, are not implemented. 
In the NCA the caching scheme which consumes less power is the one in which each SBS stores the most probable files.

 Moving from the static to the mobile transmitters scenario the minimum required power increases. This is not surprising since the number of links affected by fading increases. However, we note that in the direct file probability case, the optimal cache allocation in both the scenarios remains the same. A similar observation is also true in the inverse probability case. We performed other simulations for a third setup in which all links are affected by fading, which models a situation where the users are mobile, although we do not report here the plots for lack of space. Interestingly, also in this third scenario the optimal cache allocation is the same as in the other two. %\footnote{As expected the minimum power consumption in this third scenario is larger with respect to the other two due to the presence of fading also in the MBS-user links.}
 This may suggest that imperfect knowledge of channel statistics may be tolerated at least up to a certain extent. This aspect can have important practical implications but due to the lack of space we  do not delve further into this topic and leave it  for future work.  
  \begin{figure}[t!]
\centering
\includegraphics[width=.5\textwidth]{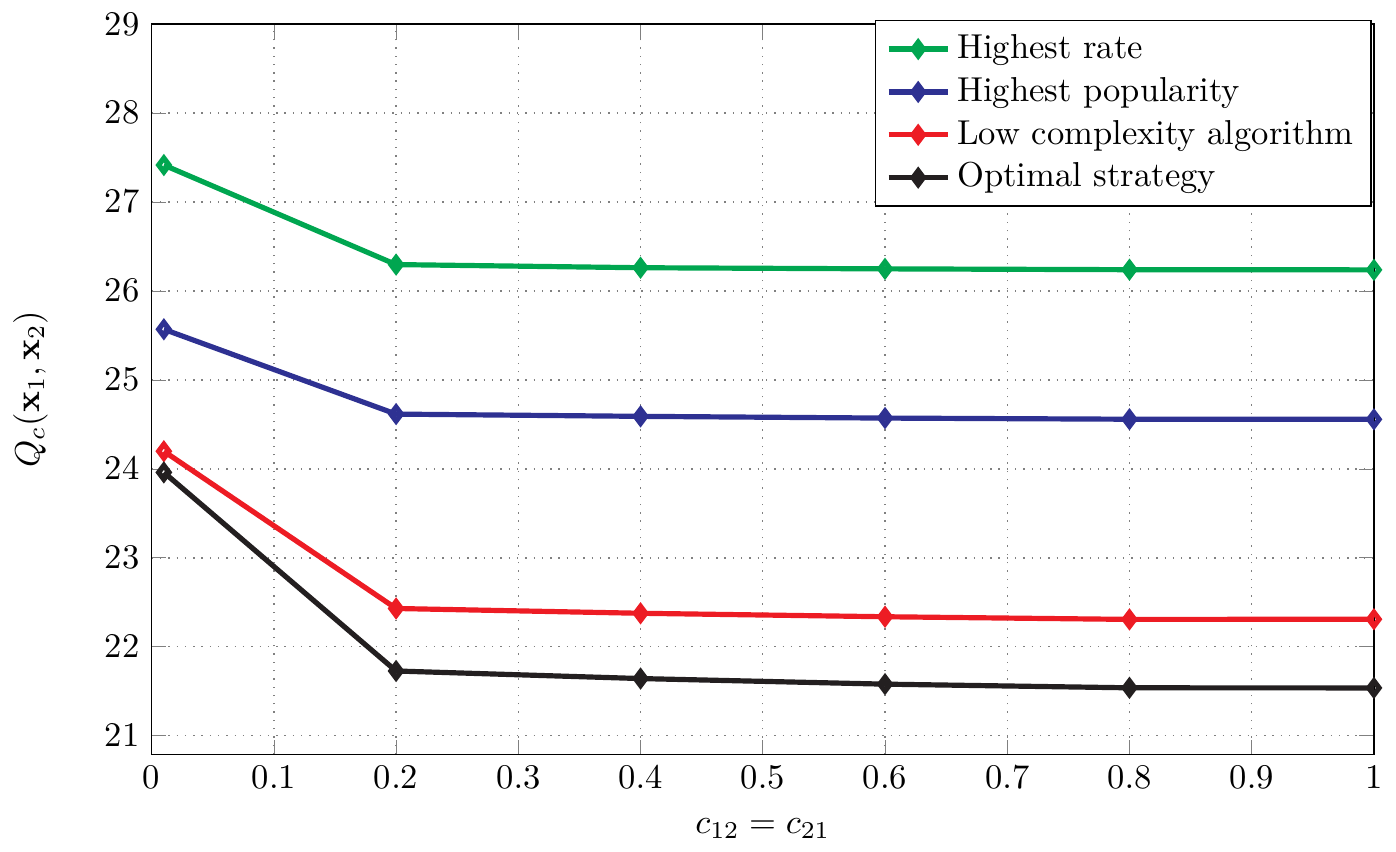}
\caption{Minimum average power cost in dB in the static scenario plotted against the interference coefficient. For $N = 10$ and $M = 2$. $R_k$ and $q_k$ for each file are described on Table \ref{tab:table2}.
The optimization is done for each average interference coefficient. In the simulations we set $a_{10} = a_{20}\ = 0.01$ and ${ a_{11} =  a_{22}\ = 1}$.}
\label{fig:AWGNrandom}
\end{figure}
 \begin{center}
\begin{table}[b] \centering
\caption{Rate and popularity ranking.}\label{tab:table2}
\begin{tabular}{llllll}
\hline
$\mathcal{F}$ &   $q_i$  &$R_i$ [bits/sec/Hz]   & $\mathcal{F}$ &   $q_i$  &$R_i$ [bits/sec/Hz]             \\
\hline
$f_1 $ & 0.5911  & 1.20 & $f_6 $ &0.0235   & 1.00   \\
$f_2 $ & 0.1697  & 0.20 & $f_7 $ &0.0178   & 1.60   \\
$f_3 $ & 0.0818  & 1.40 & $f_8 $ &0.0140   & 0.60   \\
$f_4 $ & 0.0487  & 0.80 & $f_9 $ &0.0113   & 2.00   \\
$f_5 $ & 0.0326  & 1.80 & $f_{10}$ &0.0094   & 0.40 \\
\hline 
\end{tabular}
\end{table}\end{center}
Let us compare the performance of the optimal algorithm with that of the low-complexity algorithm proposed in Section \ref{sec:imple}. Two other benchmarks are also considered, namely, one in which the files with highest popularity ranking are cached and one in which the files with the highest rate requirements are cached. The probabilities and rate of each file are given in ${\text{Table \ref{tab:table2}}}$. 
 
 We consider  $N=10$ files and a memory size for each SBS of $M=2$. The results are shown in Fig. \ref{fig:AWGNrandom} for the AWGN scenario.  The black curve represents the optimal caching strategy. The result of the proposed low-complexity algorithm is plotted in red.  As can be seen from the plot the suboptimal solution outperforms the benchmark strategies that store files with either the highest rates or the highest popularity rankings. { We recall that this is achieved with a complexity that grows linearly with $N$.} As a last remark, note that the low-complexity algorithm and the two benchmark allocate the cache independently of the average interference, while the optimal algorithm takes it into account. Although the low-complexity algorithm consumes more power than the optimal one, it has the advantage of not requiring previous knowledge of the average interference level.

\begin{figure}[t]
    \centering
    \begin{subfigure}[b]{0.21\textwidth}
        \includegraphics[width=\textwidth]{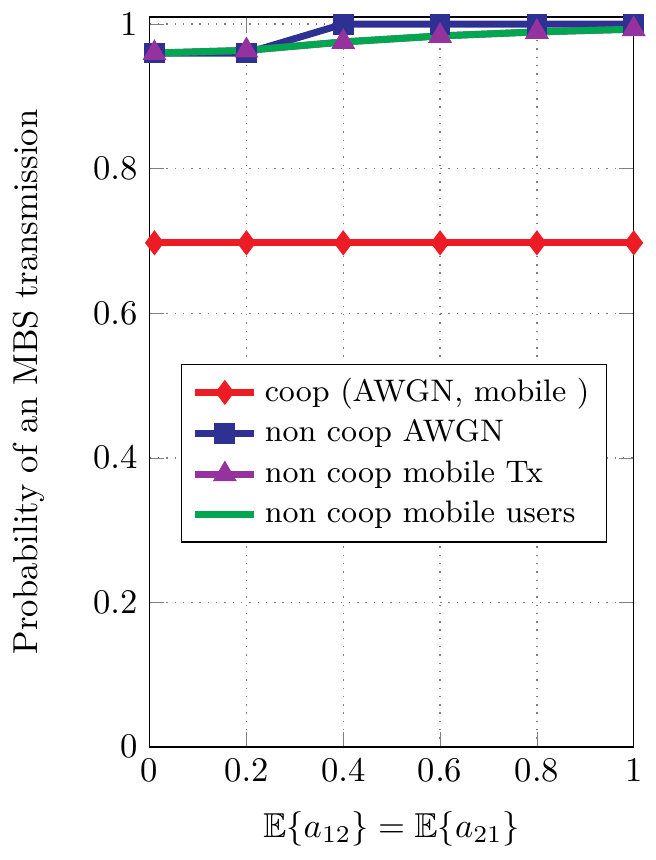}    
         \caption{Inverse}    
        \label{fig:par1}
    \end{subfigure} 
    \begin{subfigure}[b]{0.21\textwidth}
        \includegraphics[width=\textwidth]{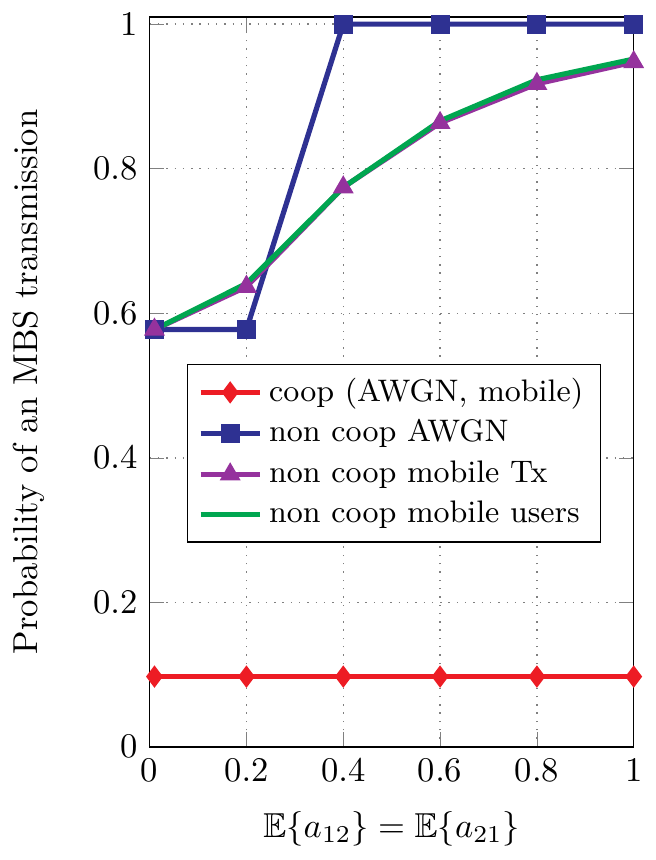}       
         \caption{Direct}
        \label{fig:par2}
    \end{subfigure}
 \small \caption{Probability of having an MBS transmission at different levels of interference for a given allocation memory. $\mathcal{M}_1=\{f_1,f_3\},   \mathcal{M}_2=\{f_2,f_4\}$  for the CA and $\mathcal{M}_1=\{f_1,f_2\}, \mathcal{M}_2=\{f_1,f_2\}$ for the NCA.}
\end{figure}

So far we saw that in some scenarios and for low levels of interference the power gain deriving from cooperation is limited. This is because we are trying to spare the usage of the MBS as much as possible. Since from the plots showed so far the MBS utilization cannot be appreciated, in  Fig. \ref{fig:par1}  and  Fig. \ref{fig:par2} we plot, for a given caching allocation and for different levels of interference, the probability that a transmission is served by the MBS for inverse and direct probabilities, respectively. For the CA we consider the cache  allocation: $\mathcal{M}_1=\{f_1,f_3\},  \mathcal{M}_2=\{f_2,f_4\}$ while for the NCA:  $\mathcal{M}_1= \mathcal{M}_2=\{f_1,f_2\}$. 

In the the CA (red curve) the probability of having a transmission from the MBS is independent from the level of interference and in the static and mobile transmitters setups the probability coincides.  This is because in the cooperative case the usage of the MBS only depends on the cache allocation of the SBSs while in the NCA it depends also on the mutual interference level.
In fact, in the non-cooperative case the probability of MBS transmissions grows with the level of interference. This is due to the fact  that, in the GIN channel, transmissions are forwarded to the MBS when a feasible solution does not exists and the probability of this happening increases with the interference level.
In the  inverse file probabilities case, using cooperation reduces the MBS usage of at least $25\%$ with respect to the non-cooperative case.
In Fig. \ref{fig:par2} we see how the usage of the MBS is greatly reduced with cooperation if files have direct probabilities. In this case, since the files stored in the SBSs' caches are those with higher popularity ranking (and rates), transmissions from the MBS  occur  with very low probability. In fact, in the CA MBS transmissions occur only in $10\%$ of the cases when  the best caching allocation is considered, with a gain of up to $90\%$ with respect to the best allocation of the non-cooperative case. However, as mentioned previously in the present section, such gain implies, in some case, a higher cost in terms of power.
 
 \section{Conclusions}\label{sec:conclusion}
 \vspace{-0.35mm}
We studied the optimal caching strategy for minimizing the power and limiting MBS use in   heterogeneous   networks  with mutual interference under different per-file rate constraints and one-shot delivery phase. We formulated and analysed the caching problem both in case of no cooperation between SBSs  as well as in the cooperative case. 
One among a set of different cooperative strategies is chosen by the transmitters depending on the cache allocation and file requests.  
We derived  for each of the considered transmission strategies the  minimum power needed for satisfying the rate constraints.
 
  %These were used as input for the cache allocation optimization problem. 
An optimal cache allocation algorithm as well as a low-complexity suboptimal algorithm were proposed. Our results show that memorizing the most popular files is not always the most convenient strategy when  rate requirements have to be respected.
 Furthermore, we showed that in the CA the probability of a transmission by the MBS only depends on the cache allocation and file popularity ranking while in the NCA such probability  also depends on the mutual interference.
Thanks to cooperation up to $90\%$ of the MBS resources can be saved with respect to the non-cooperative case. This implies a trade-off in terms of SBSs power. 
 
\vspace{-0.1in}
\section*{Appendix I}\label{sec:appendix}
 
We can rewrite the minimization problem of the the GI$_c$ as 
\begingroup
\allowdisplaybreaks
\begin{align} \label{minIC}
&\displaystyle \underset{P_i, P_j}   {\text{minimize }}  \displaystyle & & \frac{P_i^{(1)}}{a_{11} } + \frac{P_j^{(2)}}{a_{22} }, \nonumber \\
&\displaystyle \text{subject to } & & \displaystyle -\sqrt{2^{2R_i}-1}+ \sqrt{P_i^{(1)}}+ \sqrt{c_{12} \cdot P_j^{(2)}}\geq   0, \nonumber\\
&  & &-\sqrt{2^{2R_i}-1}+ \sqrt{P_j^{(2)}}+ \sqrt{c_{21} \cdot P_i^{(1)}}\geq   0, \nonumber\\
 \displaystyle & &  & P_i^{(1)}, P_j^{(2)} \geq 0 \nonumber.
\end{align}
\endgroup
Let us indicate with $\mathbf{P^*}= [P_i^*,P_j^*]$ the solution and let $L$ be the Lagrange function associated with the optimization problem  
\begingroup
\allowdisplaybreaks
\begin{align}
L(\mathbf{P},\mathbf{\lambda})& = -\frac{P_i^{(1)}}{a_{11} } - \frac{P_j^{(2)}}{a_{22} } +  \nonumber\\ & \lambda_1 \left(-\sqrt{2^{2R_i}-1} + \sqrt{P_i^{(1)}} + \sqrt{c_{12}P_j^{(2)}} \right)+ \nonumber\\ 
&  \lambda_2 \left(-\sqrt{2^{2R_i}-1} + \sqrt{P_j^{(2)}} + \sqrt{c_{21}P_i^{(1)}} \right)+   \nonumber\\
& \lambda_3 \cdot P_i^{(1)} + \lambda_4 \cdot  P_j^{(2)}. 
\end{align}
\endgroup
Let us consider the points where functions $g_1, ... , g_m$  are differentiable.
Applying the KKT conditions \cite{KKT} we have 
%\vspace{-1mm}
\begingroup
\allowdisplaybreaks
\begin{align}
\label{lpi86}\frac{\delta L(\mathbf{P}, \mathbf{\lambda})}{\delta P_i} &= - \frac{1}{a_{11} }+ \frac{\lambda_1^*}{2} \frac{1}{\sqrt{P^*_i}} +  \frac{\lambda_2^*}{2} \frac{\sqrt{c_{21}}}{\sqrt{P^*_i}}+\lambda_3 = 0,\\
\label{lpj87}\frac{\delta L(\mathbf{P}, \mathbf{\lambda})}{\delta P_j} & =-\frac{1}{a_{22} }+  \frac{\lambda_1^*}{2} \frac{\sqrt{c_{12}} }{\sqrt{P^*_i}}+ \frac{\lambda_2^*}{2}\frac{1}{\sqrt{P^*_j}}+\lambda_4=0,\\
\label{g1}g_1(P^*_i, P^*_j) &= - \sqrt{2^{2R_i}-1} + \sqrt{P_j^*} + \sqrt{c_{12} \cdot P_i^*} {\geq} 0,\\
g_2(P^*_i, P^*_j)& = -\sqrt{2^{2R_i}-1} + \sqrt{P_i^*} + \sqrt{c_{21} \cdot P_j^*} {\geq} 0,\\
g_3(P^*_i, P^*_j) &=  P^*_i {\geq} 0,\\
g_4(P^*_i, P^*_j)& =  P^*_j  {\geq} 0, \\
\label{ll2}\lambda_1^* {\geq}  0, \quad \lambda_2^* {\geq} &0,\quad  \lambda_3^* {\geq} 0, \quad  \lambda_4^*{\geq}  0,\\
\label{mag1}\lambda_1^*   g_1(P^*_i, P^*_j) &= \lambda_1^* \left( -\sqrt{2^{2R_i}-1} + \sqrt{P_j^*} + \sqrt{c_{12} P_i^*} \right) {=} 0,\\
\label{mag2}\lambda_2^*    g_2(P^*_i, P^*_j) &= \lambda_2^* \left(-\sqrt{2^{2R_i}-1} + \sqrt{P_i^*} + \sqrt{c_{21} P_j^*}\right){=}0,\\
\label{mag3}\lambda_3^*   g_3(P^*_i, P^*_j) &= \lambda_3^*\cdot P^*_i {=}0,\\
\label{mag4}\lambda_4^*   g_4(P^*_i, P^*_j) &= \lambda_4^* \cdot P^*_j{=} 0.
\end{align}
\endgroup
From (\ref{lpi86})-(\ref{lpj87}) and (\ref{mag3})-(\ref{mag4})    follows  that $\lambda_3=0$ and $\lambda_4=0$ then from (\ref{lpi86}) and (\ref{lpj87}) we find 
%\vspace{0mm}
\begin{align}
\label{pi}P^*_i &= \frac{1}{4}(\lambda_1^* \cdot a_{11}  + \lambda_2^* \cdot a_{11}  \cdot \sqrt{c_{21}})^2,  \\
\label{pj}P^*_j  &= \frac{1}{4}(\lambda_1^* \cdot a_{22}  \cdot \sqrt{c_{12}}+ \lambda_2^* \cdot a_{22} )^2. 
\end{align}
By plugging in  (\ref{mag1})-(\ref{mag2})  the results of (\ref{pi})-(\ref{pj}), we get the following system of equations 
%\vspace{-.05mm}
\begin{align}
\label{l1}\lambda_1^* \big[\lambda_1^* ( a_{11}  + a_{22}   c_{12} )  & +   \lambda_2^*(a_{11} \sqrt{c_{21}}+a_{22} \sqrt{c_{12}}) -    \\ &
    2\sqrt{2^{2R_i}-1}  \big]  =0 ,\nonumber \\
\label{l2}\lambda_2^* \big[ \lambda_1^* (a_{22}  \sqrt{c_{12}}   & + a_{11} \sqrt{c_{21}} )-   2\sqrt{2^{2R_i}-1} +  \\ & 
  \lambda_2^* (a_{22} + a_{11}  c_{21}) \big]   =0. \nonumber
\end{align}
%\lambda_3^*&\cdot \frac{(\lambda_1^* \cdot a_{11}  + \lambda_2^* \cdot a_{11}  \cdot \sqrt{c_{12}})^2}{4} =0,\\
%\label{l4}\lambda_4^*&\cdot  \frac{(\lambda_1^* \cdot a_{22}   \cdot \sqrt{c_{21}}+ \lambda_2^* \cdot a_{22} )^2}{4} =0
Now we need to find the quadruples  ($\lambda_1$, $\lambda_2$, $\lambda_3$, $\lambda_4$) which solve  the system  composed by equations (\ref{lpi86}), (\ref{lpj87}), (\ref{mag3}), (\ref{mag4}), (\ref{l1}) and (\ref{l2}). We obtain the following three solutions 
\begin{align}
S_1 &= \big \{ \frac{2\sqrt{2^{2R_i}-1}-\lambda_2(a_{11} \sqrt{c_{21}}+a_{22} \sqrt{c_{12}})}{a_{11} +c_{12}a_{22} }, \\ &\frac{2\sqrt{2^{2R_i}-1}(1-\frac{\sqrt{c_{12}}a_{22} +\sqrt{a_{21}}a_{11}}{a_{11} +a_{12}a_{22} })}{a_{22} +c_{21}a_{11} -\beta},  0, 0  \big\}, \\
S_2 &=  \left\lbrace {(2\sqrt{2^{2R_i}-1})}/{(a_{11} +a_{22}  c_{12})}, 0, 0, 0 \right\rbrace, \\
S_3 &=  \left\lbrace 0, {(2\sqrt{2^{2R_i}-1})}/{(a_{22} +c_{21}a_{11}) },0,0 \right\rbrace,   
\end{align}
where $\beta=\frac{(a_{11} \sqrt{c_{21}}+a_{22} \sqrt{c_{12}}  )(\sqrt{c_{12}}a_{22} +\sqrt{c_{21}}a_{11} )}{a_{11} +c_{12}a_{22} }$.
By plugging $S_1$, $S_2$ or $S_3$ in (\ref{g1})-(\ref{ll2}) we obtain at most three feasible points $(P^*_i, P^*_j)$.
We also need to consider  pairs of values $(P_i^{(1)}, P_j^{(2)})$ where the constraint functions are not differentiable. Such pairs are $A=(0, P_j^{(2)})$, ${B=(P_i^{(1)}, 0)}$ and $C=(0,0)$. Note that (0,0) cannot give a feasible solution due to the fact that in (\ref{minIC}) we would have $\sqrt{2^{2 \cdot R_i}-1}<0$.
Let us calculate the value of  $P_i^{(1)}$ and $P_j^{(2)}$ in A and B. In point A and with \textsf{$P_i^{(1)}=0$}  we have $\frac{1}{2}\log_2\left(1 + c_{12} \cdot P_j^{(2)}\right)\geq   R_i$ and ${\frac{1}{2}\log_2\left(1 + P_j^{(2)}\right)\geq   R_i}$.
Then:
$A = \left(0, \max \left\lbrace 2^{2R_i}-1, \frac{2^{2R_i}-1}{c_{12}} \right\rbrace \right)$ and dually  ${ B = \left( \max \left\lbrace 2^{2R_i}-1, \frac{2^{2R_i}-1}{c_{21}} \right\rbrace, 0 \right)}$.
Thus, the minimum power will be among  A, B and the feasible points $(P^*_i, P^*_j)$ obtained by inserting $S_1, S_2$ and $S_3$ in (\ref{g1})-(\ref{ll2}).
The cost of this channel is $c_{GIc}(i,j)= P_i^{(1)} + P_j^{(2)}$, where $P_i^{(1)} + P_j^{(2)}$ are derived from the  previous results.
\vspace{-.14in}
\section*{Appendix II}
 
The left hand side of  Han-Kobayashi \cite{kobayashi}  inequalities  (\ref{HK18})-(\ref{HK22})  can be written as  %\vspace{-.30mm}
\begingroup
\allowdisplaybreaks
 \begin{equation*}
 \begin{aligned}
\rho_1 =& \sigma^*_1 + I(Y_1;u_1|w_1w_2), \\
\rho_2 =& \sigma^*_2 + I(Y_2;u_2|w_1w_2), \\
\rho_{12} =& \sigma_{12} +I(Y_1;u_1|w_1w_2) +I(Y_2;u_2|w_1w_2), \\
\rho_{10} =&  2  \sigma_{1}^* + 2  I(Y_1,u_1|w_1w_2)+I(Y_2;u_2|w_1w_2)+ \\
&- \left[\sigma_1^*-I(Y_2;w_1|w_2)\right]^+ + \min\left\lbrace  I(Y_2;w_2|w_1),\right.\\
&  \left. I(Y_2;w_2) + [I(Y_2;w_1|w_2)-\sigma_1^*]^+, I(Y_1;w_2|w_1), \right.  \\
& \left. I(Y_1,w_1w_2)-\sigma_1^* \right\rbrace,
 \end{aligned}
\end{equation*}where
\begin{equation*}
\begin{aligned}
\rho_{20} =&  2  \sigma_{2}^* + 2  I(Y_2,u_2|w_1w_2)+I(Y_1;u_1|w_1w_2)+\\ & -[\sigma_2^*-I(Y_1;w_2|w_1)]^+ + \min\left\lbrace I(Y_1;w_1|w_2), \right.\\
& \left. I(Y_1;w_1) + [I(Y_1;w_2|w_1)-\sigma_2^*]^+, I(Y_2;w_1|w_1), \right. \\
& \left. I(Y_2,w_1w_2)-\sigma_1^* \right\rbrace \\
\sigma^*_1 =&  \min \left\lbrace I(Y_1;w_1|w_2),I(Y_2;w_1|u_2w_2)  \right\rbrace \\
\sigma_2^* =&  \min \left\lbrace  I(Y_2;w_1|w_1),I(Y_1;w_1|u_1w_1)\right\rbrace \\
\sigma_{12} =& \min \left\lbrace I(Y_1;w_1w_2),I(Y_2;w_1w_2);I(Y_1;w_1|w_2)+ \right. \\
 &\left.   I(Y_2;w_2|w_1), I(Y_2;w_1|w_2)+I(Y_1;w_2|w_1) \right\rbrace,
 \end{aligned}
\end{equation*}
and
\begin{equation*}
\begin{aligned}
&I(Y_1;u_1|w_1w_2) = C\left(\lambda_1P_1/(1+a_{12}\lambda_2P_2)\right), \\
&I(Y_2;u_2|w_1w_2) = C\left(\lambda_2P_2/(1+a_{21}\lambda_1P_1)\right), \\
&I(Y_1;w_1|w_2) = C\left(\bar{\lambda}_1P_1/(1+\lambda_1P_1+ a_{12}\lambda_2P_2)\right), \\
&I(Y_1;w_2|w_1) = C\left(a_{12}\bar{\lambda}_2P_2/(1+\lambda_1P_1+ a_{12}\lambda_2P_2)\right), \\
&I(Y_1;w_1w_2) = C\left((\bar{\lambda}_1P_1 \tiny{+}a_{12}\bar{\lambda}_2P_2)/(1 \tiny{+}\lambda_1P_1 \tiny{+} a_{12}\lambda_2P_2)\right), \\
&I(Y_2;w_2|w_1) = C\left(\bar{\lambda}_2P_2/(1+\lambda_2P_2+ a_{21}\lambda_1P_1)\right), \\
&I(Y_2;w_1|w_2) = C\left(a_{21}\bar{\lambda}_1P_1/(1+\lambda_2P_2+ a_{21}\lambda_1P_1)\right), \\
&I(Y_2;w_1w_2) = C\left((\bar{\lambda}_2P_2 \tiny{+}a_{21}\bar{\lambda}_1P_1)/(1 \tiny{+}\lambda_2P_2 \tiny{+} a_{21}\lambda_1P_1)\right), \\
&I(Y_1;w_1) = C\left(\bar{\lambda}_1P_1/(1+\lambda_2P_2+ a_{21}\lambda_1P_1)\right),\\
&I(Y_2;w_2) = C\left(\bar{\lambda}_2P_2/(1+\lambda_1P_1+ a_{12}\lambda_2P_2)\right),\\
&I(Y_1;w_2|u_1w_1) = C\left(a_{12}\bar{\lambda}_2P_2/(1+a_{12}\lambda_2P_2)\right), \\
&I(Y_2;w_1|u_2w_2) = C\left(a_{21}\bar{\lambda}_1P_1/(1+a_{21}\lambda_1P_1)\right). \\
\end{aligned}
\end{equation*}
\endgroup\normalsize  
while  $\lambda_n=1- \bar{\lambda}_n$ represents the splitting coefficient,  $v_n$ is the private message and $w_n$ the common message. Finally, $Y_1$ and $Y_2$ indicate the output of the channels for $u_1$ and $u_2$, respectively, while $[x]^+=\max\{0, x \}$.

\bibliographystyle{IEEEtran}

  \flushend

%\bibliographystyle{IEEEtran}
%\bibliography{references}
\end{document}